\documentclass[aps,preprint,pre,groupedaddress,floatfix]{revtex4}
\usepackage{amsmath,graphicx}

\begin{document}

\title{Optimizing the accuracy of Lattice Monte Carlo algorithms for simulating diffusion}

\author{Mykyta V. Chubynsky}
\email{chubynsky@gmail.com}
\author{Gary W. Slater}
\email{gslater@uottawa.ca}
\affiliation{
Department of Physics, University of Ottawa, 150 Louis-Pasteur, Ottawa, Ontario
K1N 6N5, Canada}

\date{\today}

\begin{abstract}

The behavior of a Lattice Monte Carlo algorithm (if it is designed correctly) must approach that of the continuum system that it is designed to
simulate as the time step and the mesh step tend to zero. However,
we show for an algorithm for unbiased particle diffusion that if one
of these two parameters remains fixed, the accuracy of the algorithm
is optimal for a certain \textit{finite} value of the other
parameter. In one dimension, the optimal algorithm with moves to the
two nearest neighbor sites reproduces the correct second and fourth
moments (and minimizes the error for the higher moments at large times) of the particle distribution and preserves the first two moments of the first-passage time distributions. In two and three dimensions, the same
level of accuracy requires simultaneous moves along two axes (``diagonal'' moves). Such
moves attempting to cross an impenetrable boundary should be
projected along the boundary, rather than simply rejected. We also treat the case of absorbing boundaries.

\end{abstract}

\maketitle

\section{Introduction}
\label{sec:intro} Particles diffusing in a liquid medium perform
random walks. In a computer simulation of this process, it is
sometimes convenient to introduce a lattice, assume that the particles can
only reside at lattice sites, and allow only certain types of moves
between the sites~\cite{fingerDLA,sinnoKMC09,bernstein,saxton96,plischke99,keller02,sahimi02,guo96,mercier00,mercier01macro,mercier01noncond,gauthier02,gauthier03,casault07,kosmidis,torres08,langowski}. Different variants of such \textit{Lattice Monte
Carlo} (LMC) algorithms are distinguished by the sets of allowed
moves and the corresponding probabilities. These are chosen
depending on the purpose of the algorithm. For instance, Metropolis
Monte Carlo~\cite{metrop} can be used to study equilibrium properties, but
is in general inadequate for dynamics, especially in a strong
external field~\cite{gauthier02}. For dynamical algorithms, another important
(and often neglected) ingredient is the time step (the amount by
which the time is incremented after each attempted LMC move)~\cite{gauthier04}.
While in principle the time step can vary during a simulation, in this paper we restrict
ourselves to algorithms where it remains constant. Moreover, we only consider the case of
\textit{unbiased} diffusion, where there is no external field or
other external influence (such as a flow) that would drive the
particles preferentially in a particular direction. Some of the
approaches used in this paper are applicable to the case of biased
diffusion as well, and this will be described in a separate
publication. We also assume that the system is uniform, i.e., the diffusion constant is the same everywhere.

As an example of a dynamical LMC algorithm, consider unbiased
one-dimensional (1D) diffusion along an infinite line, with the 1D
lattice sites placed equidistantly and numbered consecutively by
integer numbers. In the simplest approach, at each step the particle
moves left or right to a neighboring site with equal probabilities.
If for a particle that is known to be at site $i$, the probability
that it moves to site $j$ during a given time step is $p_{i\to j}$,
then
\begin{eqnarray}
p_{j\to j+1}=p_{j\to j-1}&=&1/2,\label{1Dordinary1}\\
p_{j\to l}&=&0{\rm \ for\ }l\ne j\pm 1.\label{1Dordinary2}
\end{eqnarray}
The time step $\tau$ can be determined, e.g., using a mapping onto the mean first-passage time between sites~\cite{gauthier04}, and is equal
\begin{equation}
\tau=\frac{a^2}{2D},\label{tau1D}
\end{equation}
where $a$ is the distance between the lattice sites (the lattice
constant or the mesh step) and $D$ is the diffusion constant of the
particle in the medium.

Rather than working with individual particles, one can look at the evolution of particle distributions, described by the set of the mean particle numbers at each site, $\{n_i(t)\}$, where the subscript $i$ refers to the site number. In general, the mean particle numbers after the move are given in terms of those before the move by
\begin{equation}
n_j(t+\tau)=\sum_l p_{l\to j} n_l(t),\label{master}
\end{equation}
where in principle $|l-j|$ can be allowed to be larger than unity.
Equation~(\ref{master}) is known as the \textit{master equation}.
For the particular simple algorithm described by
Eqs.~(\ref{1Dordinary1})--(\ref{tau1D}),
\begin{equation}
n_j(t+\tau)=\frac{1}{2} \left( n_{j-1}(t)+n_{j+1}(t)
\right).\label{1D}
\end{equation}
The set of equations~(\ref{1D}) for all sites can be solved numerically for given initial
conditions, and this provides, in a sense, a \textit{numerically
exact} solution of the LMC algorithm~\cite{guo99}.

On the other hand, diffusion can be studied using \textit{continuum
equations} that describe the time evolution of the particle
concentration. For example, for unbiased diffusion in 1D
\begin{equation}
\frac{\partial n(x,t)}{\partial t}=D~\frac{\partial^2
n(x,t)}{\partial x^2}.\label{cont1D}
\end{equation}
To solve such an equation numerically, one can discretize it
replacing the derivatives with differences:
\begin{equation}
\frac{n(x,t+\tau)-n(x,t)}{\tau}\approx D \times
\frac{n(x+a,t)-2n(x,t)+n(x-a,t)}{a^2}
\end{equation}
or
\begin{equation}
n(x,t+\tau)\approx n(x,t)+\frac{D\tau}{a^2} \left(
n(x+a,t)-2n(x,t)+n(x-a,t) \right).
\end{equation}
If $x=ja$, where $j$ is integer, then we get a set of equations involving only the values of $n(x,t)$ at discrete points:
\begin{equation}
n_j(t+\tau)\approx n_j(t)+\frac{D\tau}{a^2} \left(
n_{j+1}(t)-2n_j(t)+n_{j-1}(t) \right), \label{contdiscr}
\end{equation}
where $n_j(t)\equiv n(ja,t)$. Note that this coincides with
Eq.~(\ref{1D}) when $\tau$ is given by Eq.~(\ref{tau1D}). In other
words, the master equation for a LMC algorithm can be viewed as a
discretization of the continuum equation for the same diffusion
problem.

\begin{figure}
\begin{center}
\includegraphics[width=3.5in]{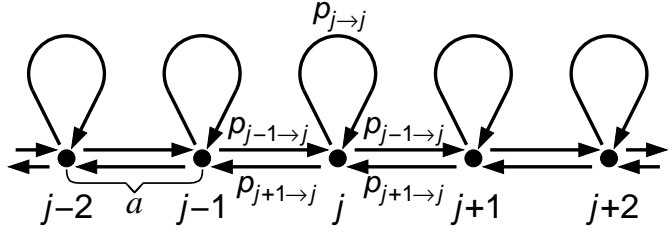}
\caption{A schematic of the 1D algorithm considered in this paper. At a particular step, in addition to the usual moves to nearest neighbors, the particle can also stay put (which can be considered as the move to the original site). The notation for the probabilities of the moves is given. These probabilities (along with the time step) can be adjusted to optimize the algorithm. For unbiased diffusion, we always have $p_{j-1\to j}=p_{j+1\to j}$.\label{fig:1Dfree}}
\end{center}
\end{figure}

Note, however, that other choices of the time step $\tau$ are possible in Eq.~(\ref{contdiscr}). For each such choice, one can define a new LMC algorithm with a new set of transition probabilities by comparing Eq.~(\ref{contdiscr}) to Eq.~(\ref{master}) (with the caveat that these probabilities have to remain bounded between zero and one). Indeed, if we rewrite Eq.~(\ref{contdiscr}) as
\begin{equation}
n_j(t+\tau)\approx \frac{D\tau}{a^2}n_{j-1}(t)+\left(1-\frac{2D\tau}{a^2}\right)n_j(t)+\frac{D\tau}{a^2}n_{j+1}(t)
\end{equation}
and compare this expression to Eq.~(\ref{master}), we obtain
\begin{eqnarray}
p_{j-1\to j}=p_{j+1\to j}&=&\frac{D\tau}{a^2},\label{p1D}\\
p_{j\to j}&=&1-\frac{2D\tau}{a^2}.\label{p01D}
\end{eqnarray}
To ensure that these probabilities remain between zero and one, the time step must be in the range
\begin{equation}
0<\tau\le \frac{a^2}{2D} .\label{ineq}
\end{equation}
Note that in general we now have a nonzero probability $p_{j\to j}$ of staying at the same site during a particular time step (Fig.~\ref{fig:1Dfree}). The only exception is the particular case of the algorithm given by Eqs.~(\ref{1Dordinary1})--(\ref{tau1D}), which corresponds to the largest possible time step,
\begin{equation}
\tau_{\rm max}=\frac{a^2}{2D}.\label{tau1D2}
\end{equation}
Such an algorithm (to which we will refer as the \textit{ordinary algorithm}) is a logical choice from the point of view of the
computational efficiency of the simulation. However, choosing a
shorter time step can increase the \textit{accuracy} of the LMC
algorithm.

The problem of optimizing the accuracy of the algorithm given by
Eqs.~(\ref{p1D}) and (\ref{p01D}) seems trivial at first. Indeed,
the smaller the time step and the mesh step, the more accurately the
discretization (\ref{contdiscr}) represents the continuum equation
(\ref{cont1D}) and thus the more accurate the corresponding LMC
algorithm. In practice, however, computational resources are limited
and one can only choose the smallest time step for which the
simulation is still feasible. The practically relevant question then
is: given that time step, what is the optimal mesh step? Based on
Eq.~(\ref{ineq}), it cannot be infinitely small. The smallest
possible mesh step (the finest space discretization), $a_{\rm
min}=\sqrt{2D\tau}$, corresponds to the ordinary algorithm described by Eqs.~(\ref{1Dordinary1})--(\ref{tau1D}).
While at first glance it seems it should provide the best accuracy,
we will see that this is not the case. Conversely, in some cases one
may wish to fix the mesh step (for example, choosing the largest
value that ensures that spatial inhomogeneities are still reproduced
accurately) and then optimize the time step. The largest possible
time step, $\tau=a^2/2D$ [again, corresponding to
Eqs.~(\ref{1Dordinary1}) and (\ref{1Dordinary2})], as mentioned, is
optimal in terms of computational speed. As for the accuracy, at
first glance it may seem that the smaller $\tau$ the better, but
again, this trivial answer turns out to be incorrect.

This problem of optimizing the accuracy of LMC algorithms for
diffusion given a fixed time step or mesh
step is the subject of this paper. While in our considerations we
fix the mesh step and allow the time step and the transition
probabilities to vary, it will be clear from our derivation that fixing
the time step instead will lead to the same probabilities and the
same relation between the mesh step and the time step at optimality, at least as long as all inhomogeneities, such as obstacles and their features, are much larger than the mesh step. In the next section, we solve the optimization
problem for the 1D diffusion algorithm described above. We do this
using three different approaches, with the same end result. In
Secs.~\ref{sec:unb2D} and \ref{sec:unb3D}, we consider the 2D and 3D analogs of the same
problem and show that the same accuracy as in 1D requires the
introduction of simultaneous moves in at least two directions, or
\textit{diagonal moves}. In Secs.~\ref{sec:bourefl} and \ref{sec:bouabs}, we show how the LMC rules should
be modified for sites next to reflecting or absorbing boundaries. We end the
paper with a discussion of our results.

\section{One dimension}
\label{sec:unb1D}
In this section, we optimize the 1D LMC algorithm
for unbiased diffusion with the moves restricted to nearest neighbors.
We do this in three different ways. The first approach is based on the
comparison between the solutions of the continuum equation
(\ref{cont1D}) and the master equation (\ref{master}) (with only
$p_{j\pm 1\to j}$ and $p_{j\to j}$ nonzero); they should match
particularly for long-wavelength, slowly decaying modes. The other two methods compare the exact moments of the particle distribution or the first-passage time (FPT) distribution with those of
the respective distributions generated by the algorithm. Using these three
different approaches reveals more properties of the optimal
algorithm, but the resulting sets of parameters are the same. We also consider the \textit{full} FPT distributions and show numerically that while for both the ordinary algorithm [Eqs.~(\ref{1Dordinary1})--(\ref{tau1D})] and the optimally accurate algorithm the distribution eventually approaches the correct one as the mesh gets finer, the latter algorithm achieves the same accuracy for a much coarser mesh.

\subsection{Comparison to the continuum equation}
\label{sec:compcont}
Note first that the general solution of the continuum equation (\ref{cont1D}) on an infinite line can be written as a sum (or, rather, an integral) over modes with different wave numbers $k$, each of which decays exponentially in time:
\begin{equation}
n(x,t)=\int_{-\infty}^{\infty} C(k)\exp \left( ikx-\alpha_c(k)t
\right) dk,\label{gensolcont}
\end{equation}
where $C(k)$ is an arbitrary complex function [except for the fact
that we need $C(-k)=C^*(k)$ for the solution to be real] and the
dispersion relation is
\begin{equation}
\alpha_c(k)=Dk^2.\label{alpha1Dcont}
\end{equation}
The general form of the master equation for a LMC algorithm in 1D where only moves between neighboring sites are allowed is
\begin{equation}
n_j(t+\tau)=p_{j\to j}n_j(t)+p_{j-1\to j}n_{j-1}(t)+p_{j+1\to j}n_{j+1}(t).\label{1Ddiscr}
\end{equation}
The general solution of the system of master equations (\ref{1Ddiscr}) can be written in a form similar to Eq.~(\ref{gensolcont}), except for the fact that the integration limits are finite due to space discretization:
\begin{equation}
n_j(t)=\int_{-\pi/a}^{\pi/a} C(k)\exp \left( ikaj-\alpha_d(k)t
\right) dk.\label{gensol1D}
\end{equation}
The dispersion relation $\alpha_d(k)$ can be obtained by
substituting Eq.~(\ref{gensol1D}) in Eq.~(\ref{1Ddiscr}):
\begin{equation}
\alpha_d(k)=-\frac{1}{\tau}\ln \left( p_{j\to j}+p_{j-1\to
j}\exp(-ika)+p_{j+1\to j}\exp(ika) \right) .\label{alpha1Ddiscr}
\end{equation}
Note that Eq.~(\ref{alpha1Ddiscr}) is the \textit{exact} dispersion
relation for the discrete equation (\ref{1Ddiscr}), even for finite
values of $a$ and $\tau$.

The closer Eq.~(\ref{alpha1Ddiscr}) approximates the dispersion relation (\ref{alpha1Dcont}) for the exact, continuum equation, the more accurate the discretization and the corresponding LMC scheme. One can argue that the behavior of the solution of the diffusion equation on the longest length scales (smallest $k$) should be reproduced in the first place, especially given that these longest-wavelength modes also take the most time to decay. For this reason, we expand the dispersion relations [Eqs.~(\ref{alpha1Dcont}) and (\ref{alpha1Ddiscr})] in powers of $k$ and try to match as many terms as possible. Since Eq.~(\ref{alpha1Dcont}) contains only a $k^2$ term, we need to find the conditions that would eliminate as many terms as possible in the series expansion of Eq.~(\ref{alpha1Ddiscr}), starting with the lowest order, except for the $k^2$ term whose coefficient should equal $D$. The $k^0$ term disappears if
\begin{equation}
p_{j\to j}+p_{j-1\to j}+p_{j+1\to j}=1.\label{1Dk0}
\end{equation}
In other words, the probabilities must be normalized. Given Eq.~(\ref{1Dk0}), we find that the $k^1$ term in Eq.~(\ref{alpha1Ddiscr}) has a zero coefficient if
\begin{equation}
\frac{ia}{\tau}(p_{j-1\to j}-p_{j+1\to j})=0.\label{1Dk1}
\end{equation}
This means that the probabilities of moving to the left and right should be equal, which makes sense since our diffusion process is unbiased. Taking conditions (\ref{1Dk0}) and (\ref{1Dk1}) into account, Eq.~(\ref{alpha1Ddiscr}) becomes
\begin{equation}
\alpha_d(k)=-\frac{1}{\tau}\ln\left(1+p_{j-1\to j}(-k^2 a^2+k^4 a^4/12-k^6a^6/360+\ldots)\right).\label{alpha1Ddiscr1}
\end{equation}
After expanding the logarithm in a series, the $k^2$ term of the resulting expression matches Eq.~(\ref{alpha1Dcont}) if
\begin{equation}
\frac{a^2p_{j-1\to j}}{\tau}=D.\label{1Dk2}
\end{equation}
This is equivalent to Eq.~(\ref{p1D}) and ensures the correct diffusion rate. Requiring that the next ($k^4$) term be equal to zero gives the expression
\begin{equation}
-\frac{1}{\tau}\left(-\frac{p_{j-1\to j}^2 a^4}{2}+\frac{p_{j-1\to j} a^4}{12}\right)=0.\label{1Dk4}
\end{equation}
The solution of this equation is simply
\begin{equation}
p_{j-1\to j}=\frac{1}{6}.\label{1Dopt0}
\end{equation}
Note that $p_{j-1\to j}=0$ is not a solution, since in the limit
$p_{j-1\to j}\to 0$ not just the expression in the parentheses in
Eq.~(\ref{1Dk4}), but also $\tau$ approaches zero and the ratio
$p_{j-1\to j}/\tau$ remains finite, according to Eq.~(\ref{1Dk2}).
This means that the $k^4$ term is not eliminated in the limit
$\tau\to 0$ and so, perhaps surprisingly, the algorithm is not
optimal in this limit. From
Eqs.~(\ref{1Dk0}), (\ref{1Dk1}), (\ref{1Dk2}) and (\ref{1Dopt0}), we
then get
\begin{eqnarray}
p_{j\pm 1\to j}&=&\frac{1}{6},\label{1Dopt1}\\
p_{j\to j}&=&\frac{2}{3},\label{1Dopt2}\\
\tau&=&\frac{a^2}{6D}.\label{1Dopt3}
\end{eqnarray}
Interestingly, the optimal algorithm (in terms of accuracy) requires
a time step that is 1/3 of the maximum value $\tau_{\rm max}$
allowed for a particular step length $a$ (as given by
Eq.~(\ref{tau1D2})); as a consequence, the particle must stay put
2/3 of the time. While this is costly in terms of computing time, it
is obviously better than the na\"{\i}ve expectation that optimality
would be achieved for $\tau\to 0$ even when $a$ is finite.

Ideally, further terms in the series expansion of
Eq.~(\ref{alpha1Ddiscr}) should also be zero. All odd-order terms are automatically zero, but for even-order terms starting with $O(k^6)$ this is
impossible to achieve since we have run out of adjustable LMC
parameters. Indeed, using Eq.~(\ref{alpha1Ddiscr1}) with $p_{j-1\to
j}=1/6$, one can check that the $k^6$ term is nonzero. Of course,
this term (and all subsequent terms) do vanish in the limit $a\to 0$
(which also automatically means $\tau\to 0$). For a fixed and finite
mesh step $a$, one can introduce more parameters by allowing longer-range
moves, but this can cause problems since it would effectively
increase the coarseness of the discretization of space (e.g., with
jumps of size $2a$, a particle could move through an obstacle of
size $a$).

\begin{figure}
\begin{center}
\includegraphics[width=3.5in]{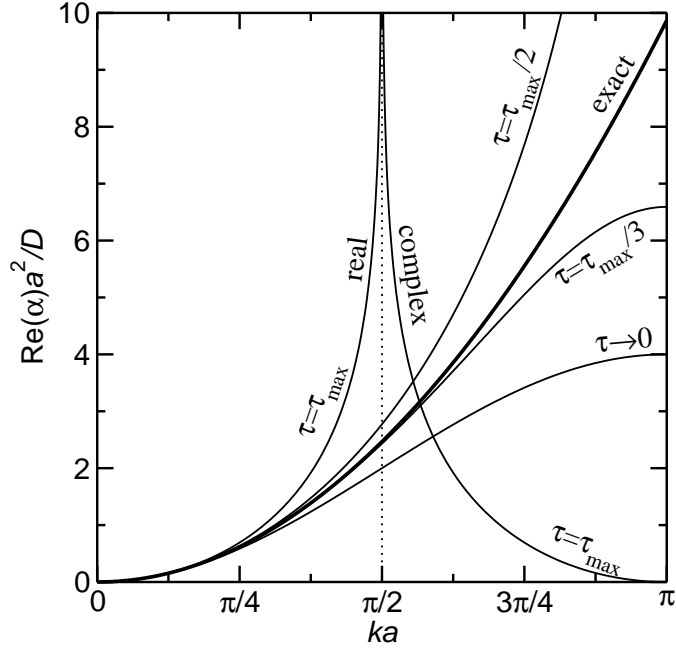}
\caption{For Fourier components of a one-dimensional particle distribution, characterized by their wavenumber $k$, a comparison between the exact decay rate obtained by solving the continuum diffusion equation [given
 by Eq.~(\ref{alpha1Dcont}); thick line] and the decay rate observed in a LMC algorithm with time step $\tau$ [given by Eq.~(\ref{alpha1Ddiscr})], for several different values of $\tau$ (thin lines). For the maximum possible time step, $\tau=\tau_{\rm max}=a^2/2D$, the decay rate diverges at $ka=\pi/2$, and for larger $ka$ it is complex becoming purely imaginary (${\rm Re}\ \alpha=0$) at $ka=\pi$. As $\tau$ decreases, the wavenumber at which the divergence occurs shifts to higher $k$ reaching $ka=\pi$ when $\tau=\tau_{\rm max}/2$. The closest matching between the exact and LMC decay rates at small $k$ is observed when $\tau=\tau_{\rm max}/3$.\label{fig:disp1D}}
\end{center}
\end{figure}

In Fig.~\ref{fig:disp1D}, we compare the exact dispersion relation
$\alpha_c(k)$, as given by Eq.~(\ref{alpha1Dcont}), with the
dispersion relations that correspond to LMC algorithms built with
different time steps $\tau$. The latter dispersion relations are
obtained using Eq.~(\ref{alpha1Ddiscr}) with the probabilities given by Eqs.~(\ref{p1D}) and (\ref{p01D}) and thus
satisfying Eqs.~(\ref{1Dk0}), (\ref{1Dk1}) and (\ref{1Dk2}), which
guarantees the correct expansion terms up to $k^2$, but the $k^4$
term only vanishes for $\tau=\frac{1}{3}\tau_{\rm max}$. For the
maximum possible value $\tau=\tau_{\rm max}=a^2/2D$, which
corresponds to the ordinary algorithm with zero probability of
staying on the site, two peculiarities catch the eye. First, at
$k=\pi/2a$, $\alpha_d(k)$ diverges, i.e., the corresponding mode
decays infinitely fast. Indeed, the distribution
\begin{equation}
\ldots,0,\frac{1}{2},1,\frac{1}{2},0,\frac{1}{2},1,\frac{1}{2},0,\ldots
\end{equation}
decays to the uniform distribution
$\ldots,\frac{1}{2},\frac{1}{2},\frac{1}{2},\ldots$ in a single time
step and does not change afterwards. Even more strikingly, for
$k=\pi/a$, the real part of $\alpha_d(k)$ is zero, i.e., this mode
does not decay at all; at the same time, the imaginary part is ${\rm
Im\ }\alpha_d(\pi/a)=\pi/\tau$. As a result, the initial
distribution
\begin{equation}
\ldots,0,1,0,1,\ldots
\end{equation}
oscillates indefinitely:
\begin{equation}
\ldots,0,1,0,1,\ldots\to\ldots,1,0,1,0,\ldots\to\ldots,0,1,0,1,\ldots\to\ldots
\label{oscil}
\end{equation}
One practical consequence of this is that a particle starting at one
of the even-numbered sites will always visit only odd-numbered sites
at odd-numbered time steps, and only even-numbered sites
at even-numbered time steps. When a smaller time step is used with the
same lattice constant $a$, there is some probability of staying on
the same site at each time step, and oscillations similar to
Eq.~(\ref{oscil}) decay. Eventually, for $\tau<\frac{1}{2}\tau_{\rm
max}$, the oscillations do not occur at all, as $\alpha_d$ is always
real in that case. The exact dispersion curve is matched best for
small $k$ when $\tau=\frac{1}{3}\tau_{\rm max}$, as expected.

Instead of using the explicit solution of the   master equations [Eq.~(\ref{gensol1D})], another possible approach is to insert the partial solutions of the continuum problem [i.e., the integrand of Eq.~(\ref{gensolcont})] in the master equation (\ref{1Ddiscr}) replacing $x$ with $ja$ and find for what values of the parameters the resulting equality is satisfied most accurately for small $k$. After the substitution, dividing both sides by $C(k)\exp(ikaj-Dk^2t)$, we obtain
\begin{equation}
\exp(-Dk^2\tau)\overset{?}{\simeq}p_{j\to j}+p_{j-1\to j}\exp(-ika)+p_{j+1\to j}\exp(ika).\label{insertion}
\end{equation}
Here we use the $\overset{?}{\simeq}$ sign as a reminder that this is a relation that needs to be verified (hence the question mark); moreover, it is not expected to hold exactly, since lattice diffusion is only an approximate representation of the continuum diffusion process (thus ``$\simeq$'').
By expanding both sides in the Taylor series in $k$ we can verify that this equality is satisfied to $O(k^4)$ only when the probabilities and the time step are given by Eqs.~(\ref{1Dopt1})--(\ref{1Dopt3}).

An algorithm with a nonzero probability to stay put was proposed before~\cite{gauthier04} for \textit{biased} diffusion in an external field. It was shown that in that case this nonzero probability was necessary even to reproduce the correct diffusion constant. Interestingly, the zero-field limit of that algorithm coincides with the optimal algorithm derived here, even though in zero field even the ordinary algorithm reproduces the diffusion constant correctly.

\subsection{Moments of the particle distribution}
\label{sec:moms}
Another way of constructing and analyzing LMC algorithms is by
looking at the moments of the particle distribution. In continuum
space, for particles starting at the origin ($x=0$) at $t=0$, the
distribution at time $t$ is
\begin{equation}
n(x,t)=\frac{1}{2\sqrt{\pi Dt}}e^{-x^2/4Dt}.
\end{equation}
The moments of this distribution are
\begin{equation}
\langle x^{2m}\rangle =(2m-1)!! (2Dt)^m = (2m-1)!! \times \langle x^{2}\rangle^m.
\end{equation}
All odd moments are zero. The first three nonzero moments are
\begin{eqnarray}
\langle x^2\rangle&=&2Dt,\label{exmom2}\\
\langle x^4\rangle&=&12D^2t^2,\label{exmom4}\\
\langle x^6\rangle&=&120D^3t^3.\label{exmom6}
\end{eqnarray}

We will now compute the moments of the particle distribution during
a LMC random walk. For a lattice walk starting at site 0, the
position after $N$ steps is
\begin{equation}
x_N=a\sum_{i=1}^N \eta_i,\label{xN}
\end{equation}
where $\eta_i$ is $+1$ ($-1$) for a move to the right (left) and 0 when the particle does not move. We assume from the outset that $p_{j-1\to j}=p_{j+1\to j}$, which makes all odd moments zero automatically. Before proceeding, it is convenient to note that for any positive integer $m$,
\begin{eqnarray}
\langle\eta^{2m}\rangle &=&0^{2m}\times p_{j\to j}+1^{2m}\times p_{j-1\to j}+(-1)^{2m}\times p_{j+1\to j}\nonumber\\
&=&p_{j-1\to j}+p_{j+1\to j}=2p_{j-1\to j},\\
\langle\eta^{2m-1}\rangle &=&0^{2m-1}\times p_{j\to j}+1^{2m-1}\times p_{j-1\to j}+(-1)^{2m-1}\times p_{j+1\to j}\nonumber\\
&=&p_{j-1\to j}-p_{j+1\to j}=0.
\end{eqnarray}
The second moment, which is the average square of Eq.~(\ref{xN}), is
\begin{equation}
\langle x_N^2\rangle=a^2\sum_{i,k=1}^N\langle \eta_i \eta_k \rangle.
\end{equation}
Since different steps are uncorrelated,
\begin{equation}
\langle \eta_i \eta_k \rangle=\delta_{ik}\langle \eta^2 \rangle
=2\delta_{ik}p_{j-1\to j}.
\end{equation}
Here and below $\delta_{i_1\ldots i_m}$ is (the generalization of) the Kronecker's delta, which is unity when all indices coincide and zero otherwise. We obtain
\begin{equation}
\langle x_N^2\rangle=2a^2p_{j-1\to j}\sum_{i,k=1}^N \delta_{ik}=2a^2p_{j-1\to j}N=\frac{2a^2p_{j-1\to j}}{\tau}t.
\end{equation}
This coincides with the continuum result, Eq.~(\ref{exmom2}), when $a^2p_{j-1\to j}/\tau=D$. This is the same as Eq.~(\ref{p1D}) [or (\ref{1Dk2})] that is obtainable from matching the $k^2$ terms in the continuum and discrete dispersion relations. Likewise,
\begin{equation}
\langle x_N^4\rangle=a^4\sum_{i,k,r,s=1}^N \langle \eta_i \eta_k \eta_r \eta_s \rangle .
\end{equation}
The average $\langle \eta_i \eta_k \eta_r \eta_s \rangle$ equals: 1) $\langle \eta^2 \rangle^2$ when there are two pairs of equal indices, but the indices in different pairs are different; 2) $\langle \eta^4\rangle$ when all four indices are equal; 3) 0 otherwise. As a single expression,
\begin{eqnarray}
\langle \eta_i \eta_k \eta_r \eta_s \rangle&=&(\delta_{ik}\delta_{rs}+\delta_{ir}\delta_{ks}+\delta_{is}\delta_{kr}-3\delta_{ikrs})\langle \eta^2\rangle^2+\delta_{ikrs}\langle \eta^4\rangle\nonumber\\
&=&4(\delta_{ik}\delta_{rs}+\delta_{ir}\delta_{ks}+\delta_{is}\delta_{kr}-3\delta_{ikrs})p_{j-1\to j}^2+2\delta_{ikrs}p_{j-1\to j},\label{eta4}
\end{eqnarray}
and we get
\begin{eqnarray}
\langle x_N^4\rangle&=&a^4[(12N^2-12N)p_{j-1\to j}^2+2Np_{j-1\to j}]\nonumber\\
&=&a^4[12(t^2/\tau^2)p_{j-1\to j}^2+(t/\tau)(2p_{j-1\to j}-12p_{j-1\to j}^2)].
\end{eqnarray}
When Eq.~(\ref{p1D}) is satisfied, so the second moment is correct, this becomes
\begin{equation}
\langle x_N^4\rangle=12D^2t^2+\frac{(2p_{j-1\to j}-12p_{j-1\to j}^2)a^4}{\tau}t.
\end{equation}
This approaches the continuum result, Eq.~(\ref{exmom4}), when
$t\to\infty$, but only coincides with it exactly at \textit{all}
times, if $p_{j-1\to j}=1/6$, which coincides with
Eq.~(\ref{1Dopt0}) and gives rise to the optimal algorithm
[Eqs.~(\ref{1Dopt1})--(\ref{1Dopt3})].

In general, even moments are given by
\begin{equation}
\langle x_N^{2m}\rangle = a^{2m}\sum_{i_j=1}^N \langle \eta_{i_1}\ldots\eta_{i_{2m}}\rangle.
\end{equation}
For the sixth moment, we need $\langle \eta_{i_1}\ldots\eta_{i_6}\rangle$. This equals: 1) $\langle \eta^2\rangle^3$ when there are three pairs of equal indices, but no equality of indices between any of the pairs; 2) $\langle \eta^2\rangle\langle \eta^4\rangle$ when there are four equal indices and another pair of equal indices not equal to the first four; 3) $\langle \eta^6\rangle$ when all six indices are equal; 4) 0 otherwise. When the indices run from 1 to $N$, there are $15N(N-1)(N-2)$ combinations of indices of the first type, $15N(N-1)$ combinations of the second type, and $N$ combinations of the third type. Then
\begin{eqnarray}
\langle x_N^6\rangle&=&a^6[15N(N-1)(N-2)\langle \eta^2\rangle^3+15N(N-1)\langle \eta^2\rangle\langle \eta^4\rangle+N\langle \eta^6\rangle]\nonumber\\
&=&a^6[120N(N-1)(N-2)p_{j-1\to j}^3+60N(N-1)p_{j-1\to j}^2+2Np_{j-1\to j}]\nonumber\\
&=&a^6[120(t/\tau)^3p_{j-1\to j}^3+(t/\tau)^2(60p_{j-1\to j}^2-360p_{j-1\to j}^3)\nonumber\\
& &\hspace{2cm}+(t/\tau)(2p_{j-1\to j}-60p_{j-1\to j}^2+240p_{j-1\to j}^3)].
\end{eqnarray}
It is easy to check that the coefficient of the $t^3$ term matches
Eq.~(\ref{exmom6}) whenever Eq.~(\ref{p1D}) is satisfied, i.e.,
whenever the second moment (and the leading term in the fourth
moment) are correct. Moreover, the $t^2$ term is zero if and only if
$p_{j-1\to j}=1/6$, i.e., for the optimal algorithm. However, even
in this case the remaining $t^1$ term is incorrect. Yet, the error of the
sixth moment at large $t$ is optimized, since in this case the
relative error is $O(t^{-2})$, whereas in all other cases the $t^2$
term is present and the error is $O(t^{-1})$.

In fact, this statement about the optimization of the error is true
for \textit{all} moments. In the case of the $(2m)$-th moment, the two terms of the highest order in $t$ are of order $t^m$ and $t^{m-1}$. By analogy with the
sixth moment, the only combinations of indices in $\langle
\eta_{i_1}\ldots\eta_{i_{2m}}\rangle$ contributing to these terms
are: 1) those with $m$ pairs of equal indices with no equality
between pairs; 2) those with four equal indices and $m-2$ pairs of
equal indices with no equality between these groups. For the first
type, there are $(2m-1)!!$ ways to break the indices into pairs and
$N!/(N-m)!$ ways to assign the values of the indices to these pairs,
for the total of
 $(2m-1)!!N!/(N-m)!$ terms in the sum, each of which
is $\langle\eta^2\rangle^m$. For the second type, there are
$\binom{2m}{4}$ ways to choose the group of four indices, $(2m-5)!!$
ways to break the remaining indices into pairs, and $N!/(N-m+1)!$
ways to assign the values of the indices to all these groups, for
the total of $\binom{2m}{4}(2m-5)!!N!/(N-m+1)!$ terms, each of which
is $\langle\eta^4\rangle\langle\eta^2\rangle^{m-2}$. The $(2m)$-th
moment then is
\begin{eqnarray}
\langle x_N^{2m}\rangle&=& a^{2m}\left[\frac{(2m-1)!!N!}{(N-m)!}\langle\eta^2\rangle^m+\frac{\binom{2m}{4}(2m-5)!!N!}{(N-m+1)!}\langle\eta^4\rangle\langle\eta^2\rangle^{m-2}+O(N^{m-2}) \right]\nonumber\\
&=&a^{2m}\left[(2m-1)!!\left(N^m-\frac{m(m-1)}{2}N^{m-1}\right)(2p_{j-1\to j})^m\right.\nonumber\\
& &\left.\hspace{1cm}+\binom{2m}{4}(2m-5)!!N^{m-1}(2p_{j-1\to j})^{m-1}+O(N^{m-2})\right]\nonumber\\
&=&a^{2m}\left[(2m-1)!!\left(\frac{2p_{j-1\to j}}{\tau}\right)^m t^m\right.\nonumber\\
& &\left.\hspace{1cm}+(2m-1)!!m(m-1)2^{m-1}\frac{p_{j-1\to j}^{m-1}/6-p_{j-1\to j}^m}{\tau^{m-1}}t^{m-1} + O(t^{m-2})\right].
\end{eqnarray}
We see again that the coefficient of the leading $t^m$ term is
correct whenever Eq.~(\ref{p1D}) is satisfied, and the subleading
term vanishes when $p_{j-1\to j}=1/6$.

\subsection{Moments of the FPT distribution}
\label{sec:FPT}
Consider again a particle diffusing in continuum space and starting at $x=0$ at time $t=0$. In the FPT problem~\cite{redner}, we consider two imaginary walls at $x=\pm b$ and find the first instance at which the particle reaches one of these walls. The corresponding time $T$ is the FPT. The LMC analog of the FPT problem would be starting at site 0 at $t=0$ and determining the first time one of the sites $\pm N$ is reached. In a good LMC algorithm, the corresponding FPT distribution should match the continuum one for $b=Na$ as closely as possible.

Consider first the case $N=1$ (and thus $b=a$). In the LMC, the FPT then corresponds to the step at which the particle first moves out of site 0 (and into one of its neighboring sites $\pm 1$). This happens at $m$th step with probability
\begin{equation}
\pi_m = p_{j\to j}^{m-1}(1-p_{j\to j}).
\end{equation}
Then the first moment of the FPT, or the mean FPT (MFPT), is
\begin{equation}
\langle T_1 \rangle_d = \tau\sum_{m=1}^{\infty}m\pi_m = \frac{\tau}{1-p_{j\to j}}=\frac{a^2}{2D},\label{Td}
\end{equation}
where Eq.~(\ref{p01D}) was used, the subscript ``1'' denotes $N=1$ and the subscript ``$d$'' stands for ``discrete.'' The second moment, or the mean-square FPT (MSFPT), is
\begin{equation}
\langle T_1^2 \rangle_d = \tau^2\sum_{m=1}^{\infty}m^2\pi_m = \frac{\tau^2 (1+p_{j\to j})}{(1-p_{j\to j})^2}=\frac{a^4}{4D^2}(1+p_{j\to
 j}).\label{T2d}
\end{equation}
On the other hand, the corresponding continuum results are~\cite{redner}
\begin{eqnarray}
\langle
 T_1 \rangle_c &=& \frac{a^2}{2D},\label{Tc}\\
\langle T_1^2 \rangle_c &=& \frac{5a^4}{12D^2}.\label{T2c}
\end{eqnarray}
Note that while the results for the MFPT are always the same ($\langle T_1 \rangle_d=\langle T_1 \rangle_c$), those for the MSFPT only coincide for $p_{j\to j}=2/3$, which corresponds to the optimal algorithm. Note also that the MSFPT for the LMC [Eq.~(\ref{T2d})] is the lowest for $p_{j\to j}=0$ (the ``trivial'' ordinary algorithm). In fact, in this case the MSFPT is simply the square of the MFPT, i.e., the variance of the FPT is zero --- the FPT is deterministic and is always equal to the time step of the algorithm. On the other hand, the optimal algorithm produces the correct variance of the FPT.

\begin{figure}
\begin{center}
\includegraphics[width=3.5in]{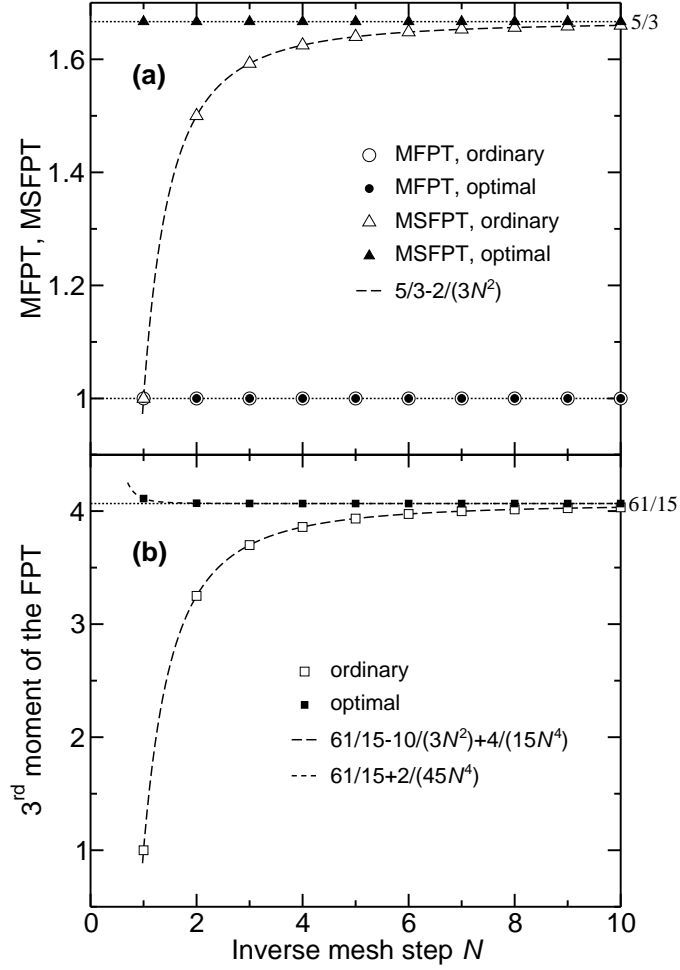}
\caption{For the ordinary [Eqs.~(\ref{1Dordinary1})--(\ref{tau1D})] and optimal [Eqs.~(\ref{1Dopt1})--(\ref{1Dopt3})] LMC algorithms, the first two moments [MFPT and MSFPT; panel (a)] and the third moment [panel (b)] of the FPT that would be observed in the simulations of 1D diffusion with $D=1/2$ starting at $x=0$ at $t=0$ and with the walls located at $x=\pm 1$ (i.e., $b=1$), as a function of the inverse mesh step $N$. The data points are obtained by numerically iterating the corresponding master equations. For the optimal algorithm, the MFPT and MSFPT are independent of $N$ and coincide with the continuum values (1 and $5/3$, respectively, shown by the dotted lines). For the ordinary algorithm, the MFPT is likewise $N$-independent, but the MSFPT data lie on the curve given by Eq.~(\ref{MSFPTord}) (dashed line). For the third moment, the dashed lines are perfect fits to the data, and the dotted line represents the continuum value.\label{fig:MFPT}}
\end{center}
\end{figure}

In fact, the matching of the first and second moments of the FPT for $N=1$ guarantees such matching for any other $N$ and even for ``asymmetric'' FPT problems where the two walls are at different distances from the initial position. This can be seen from the following consideration. Consider a random walk in continuum space and map it onto a lattice walk by introducing sites at points $x=ja$ and making a jump of the lattice walk from site $j$ to site $j\pm 1$ when the continuum walk visits site $j\pm 1$ for the first time after visiting site $j$. The probability of any particular sequence of visited sites in such a walk will be the same as in the LMC algorithms (either the ordinary or the optimal one), since the only requirement for the correct sequence is that the walk be unbiased. This means that the probability $P_m$ to reach in $m$ jumps a particular set of sites of interest starting from another specific site will always be the same for the walk obtained by direct mapping from the continuum walk and for any of the LMC approximations (where for the optimal LMC only the actual jumps are counted and not the steps where the particle stays put). On the other hand, the mean waiting times for a single jump are given by Eq.~(\ref{Td}) for the LMC algorithms and by Eq.~(\ref{Tc}) for the walk obtained by mapping from the continuum walk; likewise, the mean-square waiting times are given by Eqs.~(\ref{T2d}) and (\ref{T2c}), respectively. Therefore, if it is known that the first-passage number of jumps is $m$, then the MFPT is $m\langle T_1\rangle_{d,c}$ and the MSFPT is $m(m-1)\langle T_1\rangle_{d,c}^2+m\langle T_1^2\rangle_{d,c}$, where the appropriate subscript $d$ or $c$ and the appropriate $p_{j\to j}$ are chosen depending on the walk. After averaging over all possible $m$ we get for the MFPT
\begin{equation}
\langle T\rangle=\sum_m P_m m\langle T_1\rangle_{d,c}
\end{equation}
and for the MSFPT,
\begin{equation}
\langle T^2\rangle=\sum_m P_m (m(m-1)\langle T_1\rangle_{d,c}^2+m\langle T_1^2\rangle_{d,c}).\label{T2tot}
\end{equation}
Given that $P_m$ are the same for all three walks and since $\langle T_1\rangle_d=\langle T_1\rangle_c$ for all $p_{j\to j}$, the MFPT for both LMC algorithms is always the same as for the walk obtained by mapping the continuum walk and thus the same as for the continuum walk. This is also true for the MSFPT for the optimal LMC algorithm, as in that case $\langle T_1^2\rangle_d=\langle T_1^2\rangle_c$, but not for the ordinary LMC, as in that case this equality does not hold. Instead, for the ordinary LMC, using Eq.~(\ref{T2tot}) and Eqs.~(\ref{Td})--(\ref{T2d}) with $p_{j\to j}=0$,
\begin{equation}
\langle T^2\rangle=\sum_m P_m [m(m-1)(a^4/4D^2)+m(a^4/4D^2)]=\frac{a^4}{4D^2}\sum_m m^2P_m=\frac{a^4}{4D^2}\langle m^2\rangle,
\end{equation}
where $\langle m^2 \rangle$ is the mean-square first-passage number of jumps of the unbiased random walk. This is just the square of the (deterministic) interval between jumps (equal to $\tau$) times $\langle m^2\rangle$. For the FPT problem where the particle starts at site 0 and the walls are at sites $\pm N$, $\langle m^2\rangle$ is~\cite{weiss65}
\begin{equation}
\langle m^2\rangle=\frac{5N^4-2N^2}{3},
\end{equation}
so
\begin{equation}
\langle T^2\rangle=\frac{a^4}{12D^2}(5N^4-2N^2).
\end{equation}
If $a=b/N$, then
\begin{equation}
\langle T^2\rangle=\frac{b^4}{12D^2}(5-2/N^2),\label{MSFPTord}
\end{equation}
which coincides with the continuum result $5b^4/(12D^2)$ in the limit $N\to\infty$, but deviates for finite $N$. This is illustrated in Fig.~\ref{fig:MFPT}(a), where the comparison to the optimal algorithm is made; for the latter, as mentioned, the exact result is recovered for all $N$. Note that the data in Fig.~\ref{fig:MFPT} are obtained for $b=1$, which gives $a=1/N$, and so $N=1/a$ is simply the inverse mesh step.

In Fig.~(\ref{fig:MFPT})(b), we show the third moment of the FPT. These results have been obtained numerically, by iterating the corresponding master equations, which, of course, is much more accurate than would be achievable with actual LMC simulations. As expected, the results are now $N$-dependent in both cases, but the deviation from the asymptotic value is much smaller for the optimal algorithm and, in fact, is barely visible already for $N=1$. While we have not derived the analytical results (this would be very tedious to do), the equations given in the figure provide essentially perfect fits to the data, and we believe that these are exact (with any deviations due entirely to numerical errors). It is seen that while the deviation from the continuum value is $O(1/N^2)$ for the ordinary algorithm, it is $O(1/N^4)$ for the optimal one, in full analogy with the higher moments of the spatial particle distribution (Sec.~\ref{sec:moms}), where the leading term in the deviation from the continuum value likewise vanishes when the algorithm is optimized.

We can also obtain numerically, likewise by iterating the corresponding master equations, the whole FPT distributions for different algorithms and compare them to the continuum result. First of all, the immediate shortcoming of the ordinary algorithm is that for even $N$, the particle can only reach the boundaries at even steps, while for odd $N$, it can only reach the boundaries at odd steps. This means that the first-passage probability will vary wildly alternating between zero and a nonzero value even for arbitrarily large $N$. To eliminate this, we plot only the nonzero values of the probabilities dividing them by two, in effect, averaging between the zero and nonzero values.

\begin{figure}
\begin{center}
\includegraphics[width=3.5in]{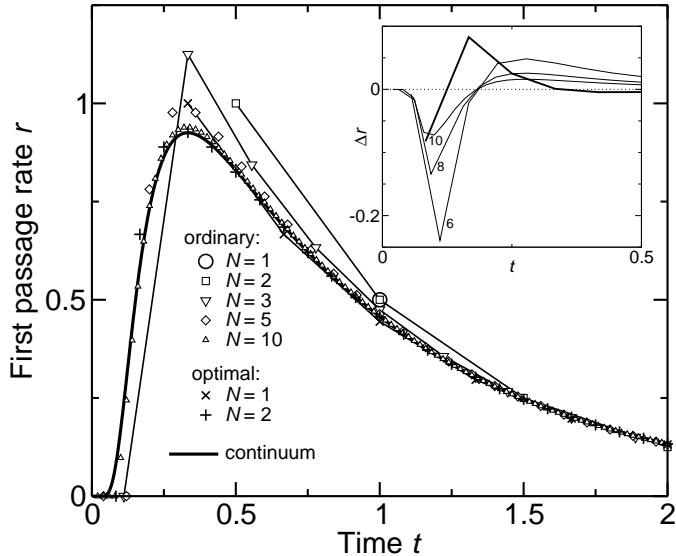}
\caption{The continuum first-passage rate for 1D diffusion with $D=1/2$ starting at $x=0$ at $t=0$ and with the walls located at $x=\pm b=\pm 1$ (thick solid line), compared to the corresponding quantity for the LMC algorithms (symbols, in some cases connected by thin lines for guidance to the eye). This first-passage rate analog is calculated as the probability of first passage at a particular step divided by the time step of the algorithm. For the ordinary algorithm, the even steps for odd $N$ and the odd steps for even $N$, at which the passage probability is always zero, are discarded and the rates for remaining time steps are divided by two. The inset shows the difference between the LMC rate and the continuum rate for the optimal algorithm with $N=2$ (thick line) and for the ordinary algorithm with $N=6$, 8, 10 (thin line), as indicated by the labels next to the curves.\label{fig:FPT}}
\end{center}
\end{figure}

The continuum result for the probability density $r(t)$ of reaching the boundary at time $t$ (or the first-passage \textit{rate}) is (see, e.g., Ref.~\cite{indian03})
\begin{equation}
r(t)=\frac{\pi D}{b^2}\sum_{m=0}^{\infty}(-1)^m (2m+1)\exp(-(2m+1)^2 \pi^2 Dt/4b^2).\label{contFPT}
\end{equation}
This is a function that is exponentially small for small $t$, reaches a maximum and decays exponentially for large $t$ (the thick solid line in Fig.~\ref{fig:FPT}).

Take for simplicity $b=1$ and $D=1/2$, which gives the MFPT $\langle T\rangle=1$.
In this case, $a=1/N$ and the time step is $1/N^2$ for the ordinary algorithm and $1/(3N^2)$ for the optimal algorithm.
The comparison of the ordinary and optimal algorithms (with different degrees of discretization $N$) with the exact result is made in Fig.~\ref{fig:FPT}. For $N=1$, the FPT is deterministic and always equals 1 for the ordinary algorithm, as the particle always moves to site $+1$ or $-1$ during the first step. For the optimal algorithm, on the other hand, the particle can move to one of the sites $\pm 1$ at any step with probability 1/3; therefore, the FPT distribution decays exponentially. While the maximum present in the continuum distribution is not reproduced, the continuum distribution is matched quite closely for large $t$. Already for $N=2$ the optimal algorithm reproduces the continuum curve very accurately. With the ordinary algorithm, a similar accuracy is not achieved until $N\approx 8$---10, as the inset of Fig.~\ref{fig:FPT} shows, where the difference between the LMC results and the continuum values is plotted. Very good results at large $t$ obtained with the optimal algorithm would not be surprising, as it is designed specifically to be as accurate as possible in this case; however, we see that it also works very well for moderate $t$. In terms of the computational effort, the ordinary algorithm with $N=8$ takes $N^2=64$ steps on average to reach the walls, whereas the optimal algorithm with $N=2$ takes $3N^2=12$ steps on average; thus the speedup is a factor of at least 5. The advantage is even larger when the ``numerically exact'' approach based on solving numerically the master equations~\cite{guo99} is used, since decreasing the mesh step increases the number of such equations and not only the number of time steps.

\begin{figure}
\begin{center}
\includegraphics[width=3.5in]{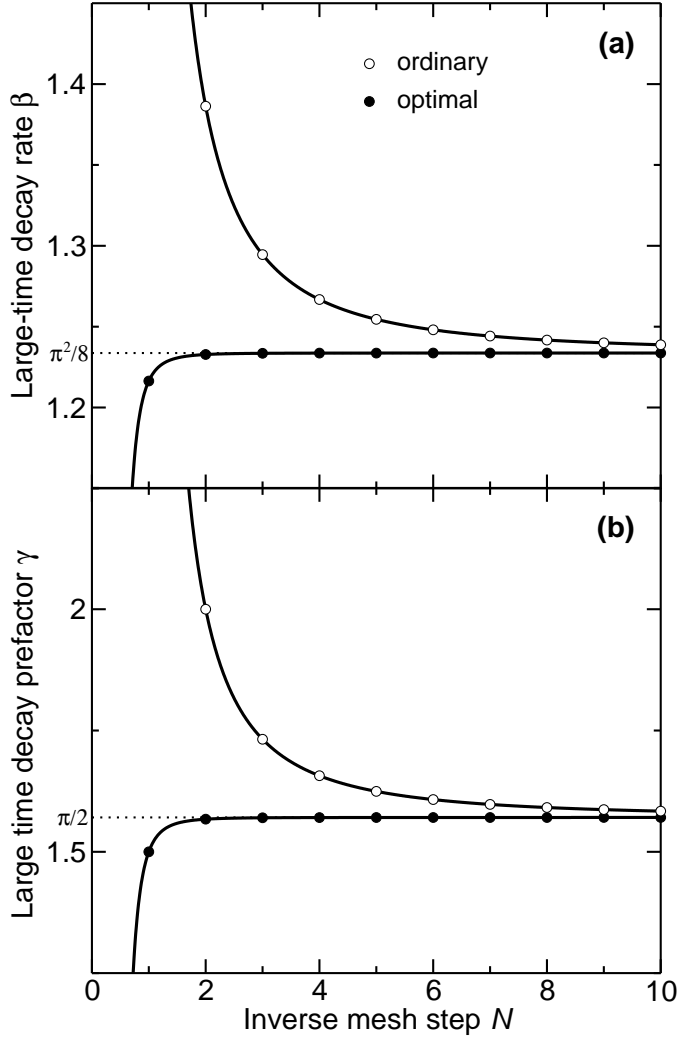}
\caption{The large-time rate of exponential decay [panel (a)] and the corresponding prefactor [panel (b)] for the first-passage rate to the walls located at $x=\pm 1$ starting at $x=0$ in simulations of 1D diffusion using the ordinary (open circles) and the optimal (filled circles) LMC algorithms. The curves are the analytical expressions [Eqs.~(\ref{betaord})--(\ref{gammaopt})]. The continuum limit values are indicated by the dotted lines.\label{fig:rate}}
\end{center}
\end{figure}

As mentioned above, the accuracy of the optimal algorithm is expected to be particularly good at large times. In the limit of large $t$, the first-passage rate decays exponentially:
\begin{equation}
r(t)\simeq \gamma\exp(-\beta t).
\end{equation}
It is therefore convenient to compare the algorithms by finding the rate of the exponential decay $\beta$ and the prefactor $\gamma$ and comparing to the continuum values. By looking at the first term of Eq.~(\ref{contFPT}), for $D=1/2$ and $b=1$ we get the continuum decay rate of $\beta_c=\pi^2/8$ and the prefactor of $\gamma_c=\pi/2$. For the LMC algorithms, these quantities can be derived as well (see Appendix). For the ordinary algorithm, the decay rate is
\begin{equation}
\beta_{\rm ord}=-N^2\ln\cos\left(\frac{\pi}{2N}\right),\label{betaord}
\end{equation}
the prefactor is
\begin{equation}
\gamma_{\rm ord}=N\tan\left(\frac{\pi}{2N}\right).
\end{equation}
For the optimal algorithm, they are respectively
\begin{equation}
\beta_{\rm opt}=-3N^2\ln\left(\frac{2}{3}+\frac{1}{3}\cos\left(\frac{\pi}{2N}\right)\right)
\end{equation}
and
\begin{equation}
\gamma_{\rm opt}=\frac{N\sin(\pi/2N)}{2/3+(1/3)\cos(\pi/2N)}.\label{gammaopt}
\end{equation}
These quantities are plotted in Fig.~\ref{fig:rate}. For the optimal algorithm, even for $N=1$ the results are quite close to the continuum value ($\beta_{\rm opt}=1.216\ldots$ vs. $\pi^2/8=1.233\ldots$ for the rate and $\gamma_{\rm opt}=1.5$ vs. $\pi/2=1.570\ldots$ for the prefactor). For the ordinary algorithm, similar accuracy is achieved for $N=4$---5; note that for $N=1$ the decay rate diverges, as the FPT is deterministic. For $N=2$ the results for the optimal algorithm are nearly indistinguishable from the continuum values on the scale of the plot and are closer than those for the ordinary algorithm with $N=10$. Note that the long-time approximation is applicable even for quite small $t$: for example, for $t=1$ (equal to the MFPT), the full sum in Eq.~(\ref{contFPT}) is $0.457365\ldots$ and its first term is $0.457436\ldots$, so the difference is only $\approx -7\times 10^{-5}$.

\section{Two dimensions}
\label{sec:unb2D} In 2D, the continuum unbiased diffusion equation
is
\begin{equation}
\frac{\partial n(x,y,t)}{\partial t}=D\left(\frac{\partial^2 n(x,y,t)}{\partial x^2}+\frac{\partial^2 n(x,y,t)}{\partial y^2}\right).\label{cont2D}
\end{equation}
Its general solution can be written in a form similar to
Eq.~(\ref{gensolcont}):
\begin{equation}
n(x,y,t)=\int_{-\infty}^{\infty}\int_{-\infty}^{\infty}
C(k_x,k_y)~\exp(ik_x x+ik_y y-\alpha_c(k_x,k_y)t)~dk_x
dk_y.\label{gensolcont2D}
\end{equation}
The dispersion relation is analogous as well:
\begin{equation}
\alpha_c(k_x,k_y)=D(k_x^2+k_y^2).\label{alpha2Dcont}
\end{equation}

\begin{figure}
\begin{center}
\includegraphics[width=3.5in]{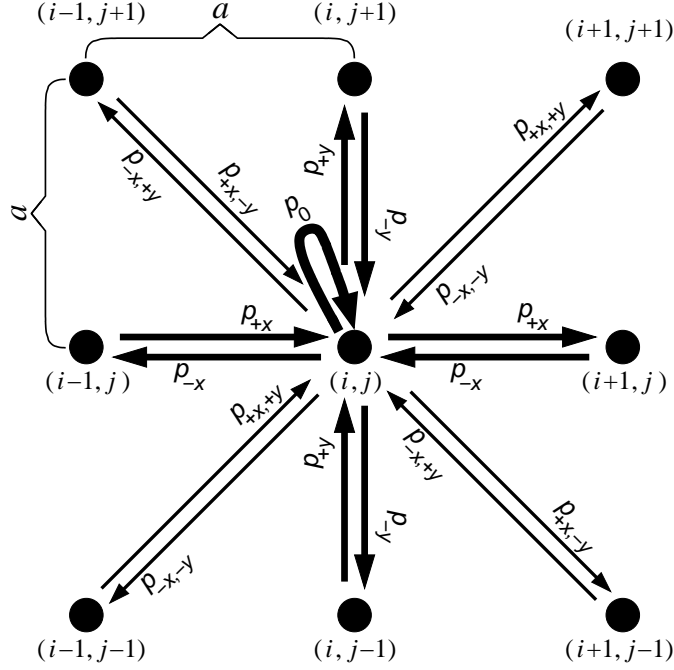}
\caption{A schematic of the general 2D LMC algorithm considered in this paper. Only moves to and from site (i,j) are shown. In addition to the traditional axial moves (arrows of medium thickness), diagonal moves (thin arrows) are introduced; also, there is a probability to stay put (represented by the thickest arrow). Notation for the probabilities of the moves is given.\label{fig:2Dfree}}
\end{center}
\end{figure}

As for the design of the corresponding LMC algorithm, we first note that different types of lattices can be chosen.
We consider the simplest choice, the square lattice, with the
lattice constant still denoted $a$. The lattice sites can be denoted
by pairs of integer numbers, $(i,j)$. If only moves along the Cartesian axes
are allowed, there are five different probabilities of moves:
$p_{(i,j)\to(i+1,j)}\equiv p_{+x}$ and $p_{(i,j)\to(i-1,j)}\equiv
p_{-x}$ for the moves along the $x$ axis in the positive and negative directions, respectively; $p_{(i,j)\to(i,j+1)}\equiv
p_{+y}$ and $p_{(i,j)\to(i,j-1)}\equiv p_{-y}$ for the moves along
the $y$ axis; and the probability of staying on the same site,
$p_{(i,j)\to(i,j)}\equiv p_0$. The master equation for the mean
particle number $n_{(j,l)}$ at a particular site $(j,l)$ is
\begin{equation}
n_{(j,l)}(t+\tau)=p_0 n_{(j,l)}(t)+p_{+x} n_{(j-1,l)}(t)+p_{-x} n_{(j+1,l)}(t)+p_{+y} n_{(j,l-1)}(t)+p_{-y} n_{(j,l+1)}(t).
\end{equation}
The general solution of the system of master equations for all sites is (again, similarly to 1D)
\begin{equation}
n_{(j,l)}(t)=\int_{-\pi/a}^{\pi/a}\int_{-\pi/a}^{\pi/a} C(k_x,k_y)\exp(ik_x aj+ik_y al-\alpha_d(k_x,k_y)t)dk\label{gensol2D}
\end{equation}
with
\begin{equation}
\alpha_d(k_x,k_y)=-\frac{1}{\tau}\ln(p_0+p_{+x}\exp(-ik_x a)+p_{-x}\exp(ik_x a)+p_{+y}\exp(-ik_y a)+p_{-y}\exp(ik_y a)).
\label{alpha2Ddiscr0}
\end{equation}
As in 1D, the goal is to choose the probabilities and the time step $\tau$ so that $\alpha_c$ and $\alpha_d$ match as closely as possible for small $k_x$ and $k_y$. While it is possible to match all coefficients of the Taylor expansion up to and including the quadratic terms (which, as in 1D, guarantees the correct diffusion rate), matching all quartic terms, like we did in 1D, is obviously impossible. This is because the form of Eq.~(\ref{alpha2Ddiscr0}) is such that there are always terms $\propto k_x^2$ and $\propto k_y^2$, but no quartic term $\propto k_x^2 k_y^2$ \textit{under the logarithm}. When the logarithm is expanded, this will necessarily produce such a quartic term. At the same time, no quartic terms are present in the expression for $\alpha_c$ [Eq.~(\ref{alpha2Dcont})] that we need to match. This reasoning is valid for \textit{any} set of moves, even if moves to second neighbors and more distant sites are present, \textit{as long as} all moves are along one of the axes. On the other hand, moves along both axes \textit{simultaneously} will produce a term $\propto k_x^2 k_y^2$ under the logarithm with a coefficient that can be chosen so that in the final expansion this term vanishes. The simplest moves of this type are the four ``diagonal'' moves whereby the particle moves by one lattice constant along the $x$ axis and simultaneously along the $y$ axis (Fig.~\ref{fig:2Dfree}). This introduces four additional probabilities: $p_{(i,j)\to(i+1,j+1)}\equiv p_{+x,+y}$, $p_{(i,j)\to(i+1,j-1)}\equiv p_{+x,-y}$, $p_{(i,j)\to(i-1,j+1)}\equiv p_{-x,+y}$, and $p_{(i,j)\to(i-1,j-1)}\equiv p_{-x,-y}$. The new master equation is
\begin{eqnarray}
n_{(j,l)}(t+\tau)=p_0 n_{(j,l)}(t)&+&p_{+x} n_{(j-1,l)}(t)+p_{-x} n_{(j+1,l)}(t)+p_{+y} n_{(j,l-1)}(t)+p_{-y} n_{(j,l+1)}(t)\nonumber\\
&+&p_{+x,+y} n_{(j-1,l-1)}(t)+p_{+x,-y} n_{(j-1,l+1)}(t)\nonumber\\
&+&p_{-x,+y} n_{(j+1,l-1)}(t)+p_{-x,-y} n_{(j+1,l+1)}(t).\label{master2D}
\end{eqnarray}
The dispersion relation now becomes
\begin{eqnarray}
\alpha_d(k_x,k_y)=-\frac{1}{\tau}\ln(p_0&+&p_{+x}\exp(-ik_x a)+p_{-x}\exp(ik_x a)+p_{+y}\exp(-ik_y a)+p_{-y}\exp(ik_y a)\nonumber\\
&+&p_{+x,+y}\exp(-ik_x a-ik_y a)+p_{+x,-y}\exp(-ik_x a+ik_y a)\nonumber\\
&+&p_{-x,+y}\exp(ik_x a-ik_y a)+p_{-x,-y}\exp(ik_x a+ik_y a)).\label{alpha2Dfull}
\end{eqnarray}
Matching the $k^0$ term,
\begin{equation}
p_0+p_{+x}+p_{-x}+p_{+y}+p_{-y}+p_{+x,+y}+p_{+x,-y}+p_{-x,+y}+p_{-x,-y}=1,
\label{2D0}
\end{equation}
which is again just the normalization statement. Matching the linear terms,
\begin{eqnarray}
p_{+x}+p_{+x,+y}+p_{+x,-y}-p_{-x}-p_{-x,+y}-p_{-x,-y}&=&0,\label{2Dlin1}\\
p_{+y}+p_{+x,+y}+p_{-x,+y}-p_{-y}-p_{+x,-y}-p_{-x,-y}&=&0.\label{2Dlin2}
\end{eqnarray}
These ensure that the average velocities are zero in both directions
(no bias). Taking Eqs.~(\ref{2D0})--(\ref{2Dlin2}) into account, we
can expand under the logarithm to obtain
\begin{eqnarray}
\alpha_d(k_x,k_y)=-\frac{1}{\tau}\ln(1&-&(C_{xx}/2) a^2 k_x^2-(C_{yy}/2) a^2 k_y^2+C_{xy} a^2 k_x k_y-i(C_{xxy}/2) a^3 k_x^2 k_y\nonumber\\
&-&i(C_{xyy}/2) a^3 k_x k_y^2+(C_{xx}/24) a^4 k_x^4+(C_{yy}/24) a^4 k_y^4-(C_{xy}/6) a^3 k_x^3 k_y\nonumber\\
&-&(C_{xy}/6) a^3 k_x k_y^3+(C_{xxyy}/4) a^4 k_x^2 k_y^2+O(k^5)),
\label{alpha2Dexp1}
\end{eqnarray}
where
\begin{eqnarray}
C_{xx}&=&p_{+x}+p_{-x}+p_{+x,+y}+p_{+x,-y}+p_{-x,+y}+p_{-x,-y}=2(p_{+x}+p_{+x,+y}+p_{+x,-y}),\label{C2D1}\\
C_{yy}&=&p_{+y}+p_{-y}+p_{+x,+y}+p_{+x,-y}+p_{-x,+y}+p_{-x,-y}=2(p_{+y}+p_{+x,+y}+p_{-x,+y}),\label{C2D2}\\
C_{xy}&=&-p_{+x,+y}+p_{+x,-y}+p_{-x,+y}-p_{-x,-y},\label{C2D3}\\
C_{xxy}&=&-p_{+x,+y}+p_{+x,-y}-p_{-x,+y}+p_{-x,-y},\label{C2D4}\\
C_{xyy}&=&-p_{+x,+y}-p_{+x,-y}+p_{-x,+y}+p_{-x,-y},\label{C2D5}\\
C_{xxyy}&=&p_{+x,+y}+p_{+x,-y}+p_{-x,+y}+p_{-x,-y}.\label{C2D6}
\end{eqnarray}
Expanding the logarithm, we now get
\begin{eqnarray}
\alpha_d(k_x,k_y)&=&\frac{1}{\tau}((C_{xx}/2) a^2 k_x^2+(C_{yy}/2) a^2 k_y^2-C_{xy} a^2 k_x k_y+i(C_{xxy}/2) a^3 k_x^2 k_y\nonumber\\
& &+i(C_{xyy}/2) a^3 k_x k_y^2+(C_{xx}^2/8-C_{xx}/24)a^4 k_x^4+(C_{yy}^2/8-C_{yy}/24)a^4 k_y^4\nonumber\\
& &+(C_{xy}/6-C_{xx}C_{xy}/2) a^3 k_x^3 k_y+(C_{xy}/6-C_{yy}C_{xy}/2) a^3 k_x k_y^3\nonumber\\
& &+(C_{xx} C_{yy}/4-C_{xxyy}/4+C_{xy}^2/2)a^4 k_x^2 k_y^2+O(k^5)).\label{alpha2Dexp2}
\end{eqnarray}
Comparing to Eq.~(\ref{alpha2Dcont}), we obtain the following
equations:
\begin{eqnarray}
C_{xx} a^2/2\tau&=&D,\label{2D1}\\
C_{yy} a^2/2\tau&=&D,\label{2D2}\\
C_{xy}/\tau&=&0,\label{2D3}\\
C_{xxy}/\tau&=&0,\label{2D4}\\
C_{xyy}/\tau&=&0,\label{2D5}\\
(C_{xx}^2/8-C_{xx}/24)/\tau&=&0,\label{2D6}\\
(C_{yy}^2/8-C_{yy}/24)/\tau&=&0,\label{2D7}\\
(C_{xy}/6-C_{xx}C_{xy}/2)/\tau&=&0,\label{2D8}\\
(C_{xy}/6-C_{yy}C_{xy}/2)/\tau&=&0,\label{2D9}\\
(C_{xx} C_{yy}/4-C_{xxyy}/4+C_{xy}^2/2)/\tau&=&0.\label{2D10}
\end{eqnarray}
From Eqs.~(\ref{2D6}) and
(\ref{2D7}), $C_{xx}=C_{yy}=\frac{1}{3}$ ($C_{xx}=0$ and $C_{yy}=0$
are not solutions, as in that case $\tau=0$). From this, using
Eq.~(\ref{2D1}) or (\ref{2D2}), we can compute the time step:
\begin{equation}
\tau=\frac{a^2}{6D}.\label{tau2D}
\end{equation}
Interestingly, this optimal time step is equal to the one we
obtained previously for the 1D problem [Eq. (\ref{1Dopt3})]. From
Eq.~(\ref{2D3}), $C_{xy}=0$. By inserting the values of $C_{xx}$,
$C_{yy}$
 and $C_{xy}$ in Eq.~(\ref{2D10}), we obtain $C_{xxyy}=1/9$.
From Eqs.~(\ref{2D4}) and (\ref{2D5}), $C_{xxy}=C_{xyy}=0$.
Equations~(\ref{2D8}) and (\ref{2D9}) are then satisfied
automatically. Thus all coefficients $C$ are known. Using
the expressions for $C$ [Eqs.~(\ref{C2D1})--(\ref{C2D6})], we obtain
linear equations for the probabilities. Together with
Eqs.~(\ref{2Dlin1}) and (\ref{2Dlin2}), this forms a linear system
of eight equations with eight unknowns that has a unique solution,
\begin{eqnarray}
p_{+x}=p_{-x}=p_{+y}=p_{-y}&=&1/9,\label{p12D}\\
p_{+x,+y}=p_{+x,-y}=p_{-x,+y}=p_{-x,-y}&=&1/36.\label{p22D}
\end{eqnarray}
Finally, from Eq.~(\ref{2D0}),
\begin{equation}
p_0=4/9.\label{p02D}
\end{equation}
The set of parameters given by Eqs. (\ref{tau2D})--(\ref{p02D}) defines an unbiased 2D LMC algorithm that is ``optimal'' for the particular set of moves considered here, since it is the only set of parameters that reproduces the continuum dispersion relation up to $O(k^5)$, the same accuracy as for the optimal 1D algorithm considered in Sec.~\ref{sec:unb1D}. [In both cases, the $k^5$ terms were not required to vanish explicitly, but are zero automatically, as are all other odd-order terms.] The set of moves is itself optimal in the sense that it is the smallest set of shortest-ranged moves that ensures such accuracy. We
note that the probability of staying put is reduced from 2/3 to 4/9,
and that the diagonal moves have small but finite probabilities.

Note that the probabilities (\ref{p12D})--(\ref{p02D}) can be obtained as products of 1D optimal probabilities [Eqs.~(\ref{1Dopt1}) and (\ref{1Dopt2})]: the probability of every 2D move is the product of the probabilities of the 1D moves that it ``consists of.'' For example, the 2D diagonal move $(+x,+y)$ can be thought of as consisting of two 1D moves, $j-1\to j$, one in the $x$ direction and one in the $y$ direction, and indeed, $p_{+x,+y}=p_{j-1\to j}\times p_{j-1\to j}$. Likewise, a move along one axis, say, $-x$, in 2D LMC can be represented as a move by one site in that direction and staying put in the orthogonal direction, which gives the correct equality $p_{-x}=p_{j+1\to j}\times p_{j\to j}$. For the probability of staying put, by similar reasoning $p_0=p_{j\to j}\times p_{j\to j}$, which is also correct. Thus the 2D optimal algorithm can be considered as the \textit{direct product} of the 1D optimal algorithm with itself, in the sense that if the 2D probabilities are arranged in a matrix, this matrix is the direct product of the vectors of the 1D probabilities:
\begin{eqnarray}
\begin{pmatrix}
p_{-x,-y} & p_{-x} & p_{-x,+y}\\
p_{-y} & p_0 & p_{+y}\\
p_{+x,-y} & p_{+x} & p_{+x,+y}
\end{pmatrix}
&=&
\begin{pmatrix}
p_{j+1\to j}\times p_{j+1\to j} & p_{j+1\to j}\times p_{j\to j} & p_{j+1\to j}\times p_{j-1\to j}\\
p_{j\to j}\times p_{j+1\to j} & p_{j\to j}\times p_{j\to j} & p_{j\to j}\times p_{j-1\to j}\\
p_{j-1\to j}\times p_{j+1\to j} & p_{j-1\to j}\times p_{j\to j} & p_{j-1\to j}\times p_{j-1\to j}
\end{pmatrix}
\nonumber\\
&=&
\begin{pmatrix}
p_{j+1\to j} \\
p_{j\to j} \\
p_{j-1\to j}
\end{pmatrix}
\otimes
\begin{pmatrix}
p_{j+1\to j} & p_{j\to j} & p_{j-1\to j}
\end{pmatrix}.
\end{eqnarray}

The fact that the direct product algorithm should work as well in 2D as its ``factors'' do in 1D can be seen directly from Eq.~(\ref{alpha2Dfull}): if the probabilities entering that equation are the products of the 1D probabilities, then the expression under the logarithm can be represented as a product of two expressions, $f(k_x)f(k_y)$, where $f(k)=p_{j\to j}+p_{j-1\to j}\exp(-ika)+p_{j+1\to j}\exp(ika)$ is the expression under the logarithm in the 1D analog of Eq.~(\ref{alpha2Dfull}), Eq.~(\ref{alpha1Ddiscr}). But if the probabilities in $f(k)$ are chosen optimally, then, according to Eq.~(\ref{alpha1Ddiscr}) and the considerations following it, $\ln(f(k))=-\tau Dk^2+O(k^6)$, therefore in Eq.~(\ref{alpha2Dfull})
\begin{equation}
\alpha_d(k_x,k_y)=-\frac{1}{\tau}\ln[f(k_x)f(k_y)]=[Dk_x^2+O(k_x^6)]+[Dk_y^2+O(k_y^6)],
\end{equation}
which matches the continuum dispersion relation [Eq.~(\ref{alpha2Dcont}] up to and including the 4th order terms. (Note, by the way, that there are no cross terms containing both $k_x$ and $k_y$ in \textit{any} order and also, as in 1D, no odd-order terms.) Note that this consideration assumes that the time step in 2D is the same as in 1D, which is indeed the case.

Moreover, the statement that the direct product algorithm is as good in 2D as its factors are in 1D is true whenever the solution of the original 2D equation can be represented as
\begin{equation}
n(x,y,t)=\int C(k_x,k_y)g_{k_x}(x,t)g_{k_y}(y,t)dk_x dk_y,
\end{equation}
where $g_k(x,t)$ are some solutions (depending on the parameter $k$) of the 1D equation for which the 1D algorithm works. This can be checked by substituting $g_{k_x}(x,t)g_{k_y}(y,t)$ into the master
equation (\ref{master2D}) [much like we did in Eq.~(\ref{insertion})], which makes both sides of that equation products of the corresponding sides of the 1D master equation (\ref{1Ddiscr}), in which $g_{k_x}$ and $g_{k_y}$ are likewise substituted. In this sense, we could have immediately ``guessed'' the correct 2D algorithm by looking at the solution of the 2D diffusion equation [Eq.~(\ref{gensolcont2D}), from which $g_{k}(x,t)=\exp(ikx-Dk^2 t)$], without the lengthy derivation that followed. That derivation was still useful, however, as it proved both the uniqueness of the solution given the set of moves and the fact that the diagonal moves were necessary.

Finally, note that from the fact that the 2D algorithm is the direct product of the 1D optimal algorithm with itself, it follows that
projecting all moves of the 2D optimal algorithm
along the $x$ axis and keeping the time step the same produces the 1D
optimal algorithm.

\section{Three and more dimensions}
\label{sec:unb3D}

In 3D, first of all, we can use the same arguments as in 2D to prove that the 3D algorithm constructed from the ``direct product'' of the 1D algorithms will reproduce the dispersion relation with the same precision, i.e., up to $O(k^5)$. This 3D algorithm involves four types of moves: in addition to moves along one of the Cartesian directions, diagonal moves along two directions and staying put, the new type of move in 3D is a move along all three directions simultaneously. The probabilities of the moves are
\begin{eqnarray}
p_0=(2/3)^3&=&8/27,\label{dirprod3D1}\\
p_{\pm x}=p_{\pm y}=p_{\pm z}=(2/3)^2\times (1/6)&=& 2/27,\\
p_{\pm x,\pm y}=p_{\pm y,\pm z}=p_{\pm x,\pm z}=(2/3)\times (1/6)^2&=& 1/54,\label{dirprod3D2}\\
p_{\pm x,\pm y,\pm z}=(1/6)^3&=& 1/216,\label{dirprod3D3}
\end{eqnarray}
and the time step is the same as in 1D and 2D:
\begin{equation}
\tau=\frac{a^2}{6D}.\label{step3D}
\end{equation}

However, unlike in 2D, the direct product algorithm is not the only way to satisfy the dispersion relation up to $O(k^5)$. For instance, there is no need to introduce moves simultaneously
along all three directions: just like the moves along two directions
are used to adjust the term $\propto k_x^2 k_y^2$, the moves along
three directions would be useful to adjust the term $\propto k_x^2
k_y^2 k_z^2$, but this term is $O(k^6)$, and the terms of this order
cannot all be made to vanish simultaneously anyway, since there are not enough different moves (i.e., not enough free parameters). Therefore, we
can retain just the moves that we used to optimize the 2D algorithm,
i.e. those along either one or two directions, and the resulting
equations are very similar. In particular, the analogs of
Eqs.~(\ref{alpha2Dfull}), (\ref{alpha2Dexp1}) and
(\ref{alpha2Dexp2}) for $\alpha_d$ are obtained simply by adding the
analogous terms involving the $z$ axis. This is also true for the
analogs of Eqs.~(\ref{2D0}), (\ref{2Dlin1}) and (\ref{2Dlin2}), to
which a similar equation obtained from equating the coefficient of
$k_z$ to zero is added. Expressions for the coefficients $C$ in
terms of the probabilities $p$ [the analogs of
Eqs.~(\ref{C2D1})--(\ref{C2D6})] as well as those for the new
coefficients involving the $z$ axis ($C_{zz}$, $C_{xz}$, $C_{yz}$,
$C_{xxz}$, $C_{yyz}$, $C_{xzz}$, $C_{yzz}$, $C_{xxzz}$, $C_{yyzz}$)
are built by analogy as well. The equations for $C$
[Eqs.~(\ref{2D1})--(\ref{2D10})] are still valid, and the analogous
equations involving the $z$ axis are added to them. From these
equations, $C_{xx}=C_{yy}=C_{zz}=1/3$,
$C_{xxyy}=C_{xxzz}=C_{yyzz}=1/9$ and all other $C$
 are zero;
the time step is again the same as in 1D and 2D, as well as in the direct product algorithm in 3D, i.e., it is given by Eq.~(\ref{step3D}). Finally, for the probabilities we obtain:
\begin{eqnarray}
p_{+x}=p_{-x}=p_{+y}=p_{-y}=p_{+z}=p_{-z}&=&1/18,\label{no3dir1}\\
p_{+x,+y}=p_{+x,-y}=p_{-x,+y}=p_{-x,-y}=p_{+x,+z}=p_{+x,-z}&=&\nonumber\\
p_{-x,+z}=p_{-x,-z}=p_{+y,+z}=p_{+y,-z}=p_{-y,+z}=p_{-y,-z}&=&1/36,\\
p_0&=&1/3.\label{no3dir3}
\end{eqnarray}
Projecting the moves of either this algorithm or the direct product algorithm [Eqs.~(\ref{dirprod3D1})--(\ref{step3D})] onto the $x$ axis produces the 1D optimal
unbiased algorithm and projecting onto the $xy$ plane produces the
2D optimal unbiased algorithm.

We now have two different ``optimal'' LMC algorithms for 3D. The second algorithm is the only possible one without moves in three directions; however, if these moves are allowed, the first algorithm is but one example. This ``direct product'' algorithm, in a way, has an advantage that \textit{some} of the sixth- and even higher-order terms in the dispersion relation vanish; in fact, it follows from a consideration analogous to that of the previous section that \textit{all} cross terms containing at least two of the three variables $k_x$, $k_y$ and $k_z$ are zero \textit{in all orders}. However, since other sixth-order terms ($k_x^6$, $k_y^6$ and $k_z^6$) do not vanish and given that, generally speaking, all sixth-order terms are expected to be of about the same magnitude, there is no guarantee that the ``direct product'' algorithm is more accurate than either the algorithm without moves in three directions or other possible algorithms where such moves are allowed. This may be problem-dependent. Note that fifth-order terms (as well as all other odd-order terms, in fact, even those odd in any one of the variables) vanish in both algorithms. On the other hand, the advantage of the algorithm without moves in three directions is that it is somewhat more short-ranged and slightly easier to program, as there are fewer moves.

We note that in the 3D case, the time step of either of the two ``optimal'' algorithms we have considered, $\tau=a^2/6D$, is also the time step
of the non-optimal but straightforward ``ordinary'' algorithm where the particle
moves at each step along one of the six Cartesian directions. This
is unlike in 1D and 2D, where the analogous straightforward
algorithms have a larger time step and thus are more computationally
efficient (as we discussed in detail for 1D). Note, however, that compared to the straightforward algorithm, we now use an extended set of moves including longer-range diagonal moves that allow an even larger time step when optimal accuracy is not required. That a larger time step is possible with the expanded set of moves is particularly obvious in 2D, where the algorithm with diagonal moves \textit{only} is equivalent to the straightforward algorithm with the mesh step of $\sqrt{2}$ times the original mesh step (and the axes rotated by $45^{\circ}$) and thus the corresponding time step is twice that of the original straightforward algorithm.

Extending these results to a space of arbitrary dimensionality using a hypercubic lattice is straightforward. First of all, ``direct product'' algorithms can be constructed, with the same properties as in 2D and 3D. On the other hand, the argument why simultaneous moves in more than two directions are unnecessary still applies (however, avoiding such moves may be impossible for other reasons, as we explain below). To construct an algorithm with moves in one direction (axial moves) and two directions (diagonal moves) only, equations analogous to Eqs.~(\ref{2D1})--(\ref{2D10}) can be written for any dimensionality, and the solutions are always the same: all coefficients $C$ with two equal indices (e.g., $C_{xx}$) are always equal to 1/3, all such coefficients with two pairs of equal indices with no equality between pairs (e.g., $C_{xxyy}$) are equal to 1/9, and all other $C$ are zero. In $d$ dimensions,
\begin{eqnarray}
C_{xx}&=&2p_1+4(d-1)p_2,\\
C_{xxyy}&=&4p_2,
\end{eqnarray}
where $p_1$ is the probability of any particular axial move and $p_2$ is the probability of any particular diagonal move. This gives
\begin{eqnarray}
p_1&=&\frac{4-d}{18},\\
p_2&=&\frac{1}{36}.
\end{eqnarray}
It is easy to check that this works for $d=1$, 2 and 3 (except for $d=1$, there are no diagonal moves and so $p_2$ is meaningless). For $d>3$, though, $p_1$ is only non-negative for $d=4$. In this case, $p_1=0$, so the optimal algorithm contains no axial moves, only diagonal ones. Since there are $\binom{4}{2}=6$ pairs of axes and thus $2\times 2\times 6=24$ different diagonal moves, the total probability of a diagonal move is $24/36=2/3$, so the probability to stay put is $p_0=1/3$, as in 3D. The time step is still $a^2/6D$. Note that this is now \textit{larger} than in the straightforward algorithm with only axial moves and zero probability to stay put; but again, a larger time step is only possible because we use longer-range diagonal moves. For $d>4$, it is impossible to make all $k^4$ terms in the dispersion relation zero with the set of moves considered. This may be possible using even longer-ranged moves (e.g., moves to second neighbors along axes) and/or simultaneous moves in three or more directions. In particular, the ``direct product'' algorithm is always an option in space of any dimensionality.

\section{Treatment of impenetrable (reflecting) boundaries}
\label{sec:bourefl}

So far, we have treated diffusion in a homogeneous
medium without any boundaries. Of course, this is a trivial
situation and the really interesting cases are those where some
spatial inhomogeneities (e.g., in the form of obstacles) are
present. In this section, we examine how to take impenetrable ``reflecting''
boundaries into account in LMC algorithms. Such boundaries are
introduced into the continuum diffusion problems via boundary
conditions, which in the unbiased case reduce to the expression
\begin{equation}
\hat{b}\cdot\overrightarrow{\nabla} n=0,\label{bcgen}
\end{equation}
where $\hat{b}$ is the unit vector normal to the boundary. Since the particle flux is proportional to the concentration gradient $\overrightarrow{\nabla} n$, Eq.~(\ref{bcgen}) ensures that there is no particle flow across the boundary.

\subsection{One dimension}
\label{sec:bou1D}
\begin{figure}
\begin{center}
\includegraphics[width=3.5in]{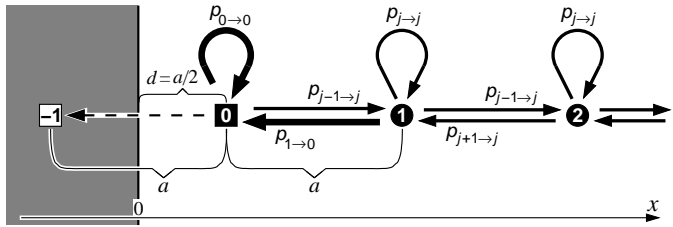}
\caption{Treatment of a reflecting boundary in the 1D optimal LMC algorithm. The inaccessible region is shaded. The site closest to the boundary is the boundary site 0 (filled square); all other sites are bulk sites (filled circles). The open square is a fictitious site in the inaccessible region useful in the analysis of the algorithm, as described in the text. The optimal placement of the boundary site is at the distance $d=a/2$ from the wall. Only the probabilities of the two moves leading to the boundary site (thick arrows), including the move from the boundary site to itself, are allowed to change (although $p_{1\to 0}$ does not in the end); the probabilities of all other moves (thin arrows) are assumed to be the same as in the bulk from the outset.  Notation for the probabilities of the moves used in the text is given. The move to the left from the boundary site (dashed arrow) is rejected and contributes to the probability of staying put.\label{fig:bou1D}}
\end{center}
\end{figure}

We consider the case of a single boundary (or wall), which means that particles cannot go beyond a certain point $x=x_b$ on the line; without the loss of generality, we choose $x_b=0$ and assume that the region $x>0$ is accessible (Fig.~\ref{fig:bou1D}). Let the site nearest to the wall have index 0 (the filled square in Fig.~\ref{fig:bou1D}). Since this site has a neighbor only on the right, moves from the left to that site are impossible, and the master equation for the mean particle number at that site has to be modified:
\begin{equation}
n_0(t+\tau)=p_{0\to 0}n_0(t)+p_{1\to 0}n_1(t).\label{1Dbound}
\end{equation}
All other sites (which we refer to as \textit{bulk sites}) have two neighbors and we assume that the corresponding master equations, including all the probabilities that are involved and the time step, remain as derived previously for free space (Fig.~\ref{fig:bou1D}). Given that assumption, the probabilities entering Eq.~(\ref{1Dbound}), $p_{0\to 0}$ and $p_{1\to 0}$, can be determined uniquely from the condition that the probabilities of all moves leaving a given site (including the move to the same site, $j\to j$) should sum up to 1 (the normalization condition). In particular, for site 0 we have
\begin{equation}
p_{0\to 0}+p_{0\to 1}=1.
\end{equation}
However, $p_{0\to 1}$ enters the master equation for site 1, which, by our assumption, remains the same as in free space, so $p_{0\to 1}$ is the same as $p_{j-1\to j}$ in free space, which determines
\begin{equation}
p_{0\to 0}=1-p_{j-1\to j}=p_{j+1\to j}+p_{j\to j}.\label{probsum}
\end{equation}
Likewise, for site 1
\begin{equation}
p_{1\to 0}+p_{1\to 1}+p_{1\to 2}=1.
\end{equation}
Here $p_{1\to 1}$ is involved in the equation for site 1 (and thus equals the ``bulk'' $p_{j\to j}$) and $p_{1\to 2}$ is involved in the equation for site 2 (thus equals the ``bulk'' $p_{j-1\to j}$), and so $p_{1\to 0}$ is determined uniquely as
\begin{equation}
p_{1\to 0}=1-p_{j\to j}-p_{j-1\to j}=p_{j+1\to j}.
\end{equation}
The time step should necessarily be the same as for bulk sites. This is because all moves out of a site must have the same time step associated with them, and for site 1, while a move $1\to 0$ leads to a boundary site and thus is considered a ``boundary'' move, a move $1\to 2$ is a ``bulk'' move. So the time step for the ``boundary'' moves is uniquely determined.

Then, for instance, for the algorithm for unbiased diffusion optimized for accuracy, Eqs.~(\ref{1Dopt1})--(\ref{1Dopt3}) lead to
\begin{eqnarray}
p_{0\to 0}&=&5/6,\label{bound1D1}\\
p_{1\to 0}&=&1/6,\label{bound1D2}\\
p_{0\to 1}&=&1/6,\\
\tau&=&a^2/6D.\label{bound1D3}
\end{eqnarray}
The probability $p_{0\to 1}$ is a bulk probability and is given here for completeness.
Note that the probability of moving to the left from site 1 is the same as from any site with index $j+1>1$ ($p_{1\to 0}=p_{j+1\to j}$). However, the probability of staying put increases by 1/6. This quantity 1/6 can be interpreted as the probability of the move to the left from site 0 that is now forbidden [in fact, this interpretation follows directly from Eq.~(\ref{probsum})]. In other words, the algorithm is: for the particle in any site, including the boundary site, select the next move according to the bulk probabilities, but if the boundary is crossed, reject the move and stay on the same site.

We can check explicitly that using Eqs.~(\ref{bound1D1})--(\ref{bound1D3}) in the master equation (\ref{1Dbound}) indeed provides the best approximation of the continuum equation. In 1D, the boundary condition [Eq.~(\ref{bcgen})] becomes
\begin{equation}
\left.\frac{\partial n(x,t)}{\partial x}\right|_{x=0}=0.\label{bc1D}
\end{equation}
The general solution of the diffusion equation [Eq.~(\ref{cont1D})] with this boundary condition can be written as
\begin{equation}
n(x,t)=\int_{0}^{\infty} C(k)\cos(kx)\exp(-Dk^2 t)dk,\label{gensol1Dbou}
\end{equation}
which is just the general solution in free space [Eq.~(\ref{gensolcont})] with the restriction $C(-k)=C(k)$ arising from the reflecting boundary condition (\ref{bc1D}) [which, together with the original condition $C(-k)=C^*(k)$, means that $C(k)$ is real].
Note that because of broken translational symmetry, \textit{generally speaking}, the solution of the discrete master equation can no longer be written in the same form we used in the bulk [Eq.~(\ref{gensol1D})]. Therefore the dispersion relation cannot be introduced and we have to use the alternative method based on inserting the integrand of the solution of the continuum equation [Eq.~(\ref{gensol1Dbou})] into the master equation and requiring that it turns into an identity as accurately as possible for small $k$ [see Eq.~(\ref{insertion}) and the accompanying discussion].

Even though the master equations for the bulk sites involve only the bulk probabilities that we have already determined in Sec.~\ref{sec:unb1D}, it is still instructive to sketch the consideration for these sites using the new form of the solution [Eq.~(\ref{gensol1Dbou})], as there are some subtleties compared to Eq.~(\ref{insertion}). Inserting the integrand of Eq.~(\ref{gensol1Dbou}) into Eq.~(\ref{1Ddiscr}) and dividing both sides by $C(k)\exp(-Dk^2t)$, we get
\begin{equation}
\cos(kx_j)\exp(-Dk^2\tau)\overset{?}{\simeq}p_{j\to j}\cos(kx_j)+p_{j-1\to j}\cos(k(x_j-a))+p_{j+1\to j}\cos(k(x_j+a)),\label{1Dbouinsert}
\end{equation}
where $x_j$ is the position of site $j$. Note, however, that expanding straightforwardly in a series in $k$ would not be correct, since $x_j$ can be arbitrarily large. We did not face this problem in Sec.~\ref{sec:unb1D}, since dividing by $\exp(ikaj)$ eliminated all dependence on $j$. However, $ka$ can still be assumed small, so the correct approach is expanding $\cos(k(x_j+a))$ in series in $ka$:
\begin{equation}
\cos(k(x_j\pm a))=\cos(kx_j)(1-k^2a^2/2+k^4a^4/24+\ldots)\mp\sin(kx_j)(ka-k^3a^3/6+\ldots).
\end{equation}
The exponential on the left-hand side of Eq.~(\ref{1Dbouinsert}) can be expanded as normal. This gives
\begin{eqnarray}
\cos(kx_j)(1-Dk^2\tau+D^2k^4\tau^2/2+\ldots)&\overset{?}{\simeq}&p_{j\to j}\cos(kx_j)\nonumber\\
& &\hspace{-3cm}+(p_{j-1\to j}+p_{j+1\to j})\cos(kx_j)(1-k^2a^2/2+k^4a^4/24+\ldots)\nonumber\\
& &\hspace{-3cm}+(p_{j-1\to j}-p_{j+1\to j})\sin(kx_j)(ka-k^3a^3/6+\ldots).
\end{eqnarray}
This should be satisfied for any $j$, thus for any $x_j$, which immediately gives $p_{j-1\to j}=p_{j+1\to j}$. Then after dividing both parts by $\cos(kx_j)$ the dependence on $x_j$ is eliminated, and we end up with a series in $k$ on both sides. Matching the coefficients gives the equations for the parameters whose only solution is given by Eqs.~(\ref{1Dopt1})--(\ref{1Dopt3}), as expected.

To obtain the probabilities for the boundary site 0, we need to repeat the same procedure for that site. Note, however, that in addition to the probabilities, there is one other free parameter: the distance $d$ from the boundary to site 0. In free space without boundaries the problem is translationally invariant, so displacement of all sites by the same amount does not influence the results, and we assumed, without any loss of generality, that site 0 had coordinate $x_0=0$. However, now that the translational symmetry of the problem is broken by the presence of the boundary, the coordinate $x_0=d$ of site 0 (Fig.~\ref{fig:bou1D}) is an important parameter. The coordinate of an arbitrary site $j$ is thus $x_j=d+aj$. Then, inserting the integrand of Eq.~(\ref{gensol1Dbou}) in Eq.~(\ref{1Dbound}) and dividing both sides by $C(k)\exp(-Dk^2t)$, we get
\begin{equation}
\cos(kd)\exp(-Dk^2\tau)\overset{?}{\simeq}p_{0\to 0}\cos(kd)+p_{1\to 0}\cos(k(d+a)).
\end{equation}
Dividing by $\cos(kd)$ and using $\tau=a^2/6D$, as in the bulk,
\begin{equation}
\exp(-a^2k^2/6)\overset{?}{\simeq}p_{0\to 0}+p_{1\to 0}\frac{\cos(k(d+a))}{\cos(kd)}.
\end{equation}
Note that $d\simeq a$, so $kd$ is small, and the Taylor expansion can be done normally, without any complications. Expanding up to $O(k^4)$,
\begin{equation}
1-a^2k^2/6+a^4k^4/72\overset{?}{=}(p_{0\to 0}+p_{1\to 0})-p_{1\to 0}[a^2k^2(1/2+d/a)-(a^4k^4/12)(1/2+2d/a-4(d/a)^3)].
\end{equation}
This equality is only satisfied when the probabilities are given by Eqs.~(\ref{bound1D1})--(\ref{bound1D3}) and
\begin{equation}
d=a/2.
\end{equation}

The meaning of the result for $d$ and the algorithm as a whole can be understood as follows. On the one hand, any solution of the continuum diffusion problem with a boundary is also a solution of the problem in free space, without a boundary, if it is continued symmetrically beyond the boundary. On the other hand, there is a similar correspondence between the discretizations of these problems as well. Indeed, looking at the corresponding master equations, the master equations for the bulk sites [Eqs.~(\ref{1Ddiscr}), (\ref{1Dopt1})--(\ref{1Dopt3})] are, of course, identical for these problems, and the equation for the boundary site [Eqs.~(\ref{1Dbound}),(\ref{bound1D1})--(\ref{bound1D3})] in the problem with the boundary is also equivalent to the equation for a bulk site, if we introduce a fictitious site (numbered $-1$) to the left of the boundary (the open square in Fig.~\ref{fig:bou1D}) and assume that the mean particle number at that site is always equal to that at the boundary site ($n_{-1}=n_0$). This matches the correspondence between the continuum problems, if the sites 0 and $-1$ are located symmetrically with respect to the boundary, which immediately gives $d=a/2$. Note also that because of the equivalence between the master equations for the boundary and bulk sites, the solution of the system of all master equations can, in fact, be represented as
\begin{equation}
n_j(t)=\int_{0}^{\pi/a} C(k)\cos(kaj)\exp \left(-\alpha_d(k)t
\right) dk.
\end{equation}
[same as Eq.~(\ref{gensol1D}) with $C(k)$ real], even though, as mentioned, this is not possible in general for an arbitrary set of parameters $p$ and $d$. Here $j$ runs from 0 to $\infty$, i.e., it includes both the boundary site and the bulk sites. The dispersion relation $\alpha_d(k)$ is, of course, exactly the same as in infinite space.

Another situation of interest is diffusion in a finite interval bounded by two reflecting walls. Since in the consideration above for a single wall the effect of that wall on the algorithm is local, only modifying the probabilities for a single boundary site, it should be possible to consider the two walls independently whenever there are at least two sites between them, with the modifications of the probabilities at each of the two boundary sites the same as in the case of a single wall. Introducing the fictitious sites behind the walls and repeating the above consideration leads to the same result. Of course, this assumes that the distance between the walls is a multiple of the mesh step $a$, so that it is possible to place both boundary sites at distance $a/2$ from the respective wall. This restricts the possible choices of $a$ given the distance between the walls.

\subsection{Two dimensions}
\label{sec:bou2D}
In 2D, the situation is more complicated, especially since diagonal moves (some of which would cross the boundary) are involved. It is no longer possible to deduce the probabilities of the moves into the boundary sites based on the normalization conditions for the probabilities alone. We restrict our consideration to the case of a perfectly flat infinite boundary parallel to one of the Cartesian axes. By analogy with 1D, the case of two parallel infinite boundaries, with the motion of the particle confined to the space between these boundaries, is easy, as each boundary can be considered independently. A more complicated and interesting case of finite and/or curved obstacles and boundaries is touched upon briefly in Sec.~\ref{sec:disc}.

\begin{figure}
\begin{center}
\includegraphics[width=3.5in]{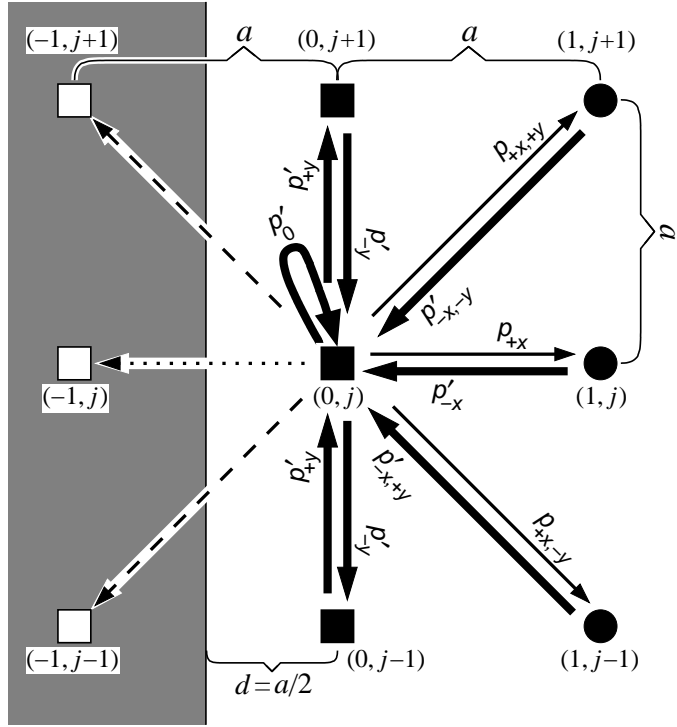}
\caption{Treatment of a reflecting boundary in the 2D optimal LMC algorithm. The diagram only shows the moves to and from one of the boundary sites [$(0,j)$]. The inaccessible region is shaded. The sites closest to the boundary are the boundary sites (filled squares); all other sites are bulk sites (filled circles). Open squares are fictitious sites in the inaccessible region useful in the analysis of the algorithm, as described in the text. The optimal placement of the line of boundary sites is at the distance $d=a/2$ from the wall. Only the probabilities of the moves to the site (thick arrows) are allowed to change compared to the bulk sites. Notation for the probabilities of the moves used in the text is given. The move to the left from the boundary site (dotted arrow) is rejected and increases the probability of staying put, but the diagonal moves across the boundary (dashed arrows) are projected along the boundary increasing $p'_{\pm y}$ compared to $p_{\pm y}$.\label{fig:bou2D}}
\end{center}
\end{figure}

Consider specifically the case of a planar boundary located at $x=0$ so that the half-space $x>0$ is accessible (Fig.~\ref{fig:bou2D}). The boundary sites form a column at $x=d$ (the filled squares in Fig.~\ref{fig:bou2D}) and have indices $(0,l)$. For a site $(j,l)$, the coordinates are $x_j=d+aj$, $y_l=al$ (there is still translational symmetry in the $y$ direction, so we can choose arbitrarily $y_0=0$). The master equations for the boundary sites are
\begin{eqnarray}
n_{(0,l)}(t+\tau)=p'_0 n_{(0,l)}(t)&+&p'_{+y} n_{(0,l-1)}(t)+p'_{-y} n_{(0,l+1)}(t)+p'_{-x} n_{(1,l)}(t)\nonumber\\
&+&p'_{-x,+y} n_{(1,l-1)}(t)+p'_{-x,-y} n_{(1,l+1)}(t),\label{master2Dbou}
\end{eqnarray}
where the primed probabilities correspond to moves ending on boundary sites and may differ from the bulk (unprimed) probabilities, whose values remain as determined in Sec.~III.

The boundary condition (\ref{bc1D}) is still true (except for the fact that the particle concentration $n$ now depends on $y$ as well), and the general solution of the diffusion equation with this boundary condition is
\begin{equation}
n(x,y,t)=\int_{k_x=0}^{\infty}\int_{k_y=-\infty}^{\infty} C(k_x,k_y)\cos(k_x x)\exp[ik_yy-D(k_x^2+k_y^2)t]dk_x dk_y.\label{gensol2Dbou}
\end{equation}
Inserting the integrand in Eq.~(\ref{master2Dbou}) and remembering that Eq.~(\ref{tau2D}) for $\tau$ should remain valid, we obtain after dividing by $C(k_x,k_y)\cos(k_xd)\exp[ik_yal-D(k_x^2+k_y^2)t]$:
\begin{eqnarray}
\exp(-a^2(k_x^2+k_y^2)/6)\overset{?}{\simeq}p'_0&+&p'_{+y}e^{-ik_ya}+p'_{-y}e^{ik_ya}\nonumber\\
&+&(p'_{-x}+p'_{-x,+y}e^{-ik_ya}+p'_{-x,-y}e^{ik_ya})
\frac{\cos(k_x(d+a))}{\cos(k_xd)}.
\label{match2Dbou}
\end{eqnarray}
After expanding up to $O(k^4)$ and equating the corresponding coefficients, we get:\bigskip\\
for $k^0$:
\begin{equation}
p'_0+p'_{+y}+p'_{-y}+p'_{-x}+p'_{-x,+y}+p'_{-x,-y}=1;\label{2Dbouk0}
\end{equation}
for $k_y^1$:
\begin{equation}
-ia(p'_{+y}-p'_{-y}+p'_{-x,+y}-p'_{-x,-y})=0;\label{2Dbouky}
\end{equation}
for $k_y^2$:
\begin{equation}
(-a^2/2)(p'_{+y}+p'_{-y}+p'_{-x,+y}+p'_{-x,-y})=-a^2/6;\label{2Dbouky2}
\end{equation}
for $k_y^3$: equivalent to Eq.~(\ref{2Dbouky});\\
for $k_y^4$: equivalent to Eq.~(\ref{2Dbouky2});\\
for $k_x^2$:
\begin{equation}
-a^2(1/2+d/a)(p'_{-x}+p'_{-x,+y}+p'_{-x,-y})=-a^2/6;\label{2Dboukx2}
\end{equation}
for $k_x^2k_y$:
\begin{equation}
ia^3(1/2+d/a)(p'_{-x,+y}-p'_{-x,-y})=0;\label{2Dboukx2ky}
\end{equation}
for $k_x^2k_y^2$:
\begin{equation}
(a^4/2)(1/2+d/a)(p'_{-x,+y}+p'_{-x,-y})=a^4/36;\label{2Dboukx2ky2}
\end{equation}
for $k_x^4$:
\begin{equation}
(a^4/12)(1/2+2d/a-4(d/a)^3)(p'_{-x}+p'_{-x,+y}+p'_{-x,-y})=a^4/72.\label{2Dboukx4}
\end{equation}
It is convenient (but not necessary) to use the normalization conditions for the probabilities. For the boundary sites,
\begin{equation}
p'_0+p'_{+y}+p'_{-y}+p_{+x,+y}+p_{+x}+p_{+x,-y}=p'_0+p'_{+y}+p'_{-y}+1/6=1;\label{2Dbousum1}
\end{equation}
for their neighbors [sites $(1,l)$],
\begin{eqnarray}
p'_{-x}+p'_{-x,+y}+p'_{-x,-y}+p_0+p_{+x}+p_{+y}+p_{-y}+p_{+x,+y}+p_{+x,-y}&=&\nonumber\\
p'_{-x}+p'_{-x,+y}+p'_{-x,-y}+5/6&=&1.\label{2Dbousum2}
\end{eqnarray}
Comparing Eqs.~(\ref{2Dbousum2}) and (\ref{2Dboukx2}), we get $d=a/2$, as in 1D, which makes Eq.~(\ref{2Dboukx4}) an identity. Then from Eqs.~(\ref{2Dboukx2ky}) and (\ref{2Dboukx2ky2}),
\begin{equation}
p'_{-x,+y}=p'_{-x,-y}=1/36\label{2Dboures1}
\end{equation}
and from Eq.~(\ref{2Dbousum2}),
\begin{equation}
p'_{-x}=1/9.\label{2Dboures2}
\end{equation}
Also using Eq.~(\ref{2Dboures1}), from Eqs.~(\ref{2Dbouky}) and (\ref{2Dbouky2})
\begin{equation}
p'_{+y}=p'_{-y}=5/36
\end{equation}
and finally, from either Eq.~(\ref{2Dbouk0}) or Eq.~(\ref{2Dbousum1}),
\begin{equation}
p'_0=5/9.\label{2Dboures3}
\end{equation}
While
\begin{eqnarray}
p'_{-x}&=&p_{-x},\label{2Dpmxbou}\\
p'_{-x,+y}&=&p_{-x,+y},\label{2Dpmxpybou}\\
p'_{-x,-y}&=&p_{-x,-y},\label{2Dpmxmybou}
\end{eqnarray}
the other three probabilities are different from the bulk values:
\begin{eqnarray}
p'_{+y}&=&p_{+y}+1/36,\label{2Dppybou}\\
p'_{-y}&=&p_{-y}+1/36,\label{2Dpmybou}\\
p'_0&=&
p_0+1/9.\label{2Dp0bou}
\end{eqnarray}
Note that not just the probability of staying put changes, so unlike in 1D, moves into the boundary are not always simply rejected. In fact, Eqs.~(\ref{2Dppybou})--(\ref{2Dp0bou}) can be rewritten as
\begin{eqnarray}
p'_{+y}&=&p_{+y}+p_{-x,+y},\label{2Dppybou1}\\
p'_{-y}&=&p_{-y}+p_{-x,-y},\label{2Dpmybou1}\\
p'_0&=&p_0+p_{-x}.\label{2Dp0bou1}
\end{eqnarray}
Therefore, the correct changes of the probabilities can be obtained, if the moves to the left $(-x)$ from the boundary sites are rejected, but the moves in the $(-x,+y)$ and $(-x,-y)$ directions are replaced by the moves in the $+y$ and $-y$ directions, respectively. In other words, it is as if the moves into the wall are replaced by their projections along the wall. We also note that, similarly to 1D, the master equations for boundary sites become equivalent to those for bulk sites when a column of fictitious sites behind the boundary is introduced (open squares in Fig.~\ref{fig:bou2D}) and the values of the mean particle numbers at these sites are made equal to those of the real boundary sites across the boundary (i.e., $n_{(-1,l)}=n_{(0,l)}$). All the probabilities could have been obtained from this consideration alone. It is also possible to obtain the 2D algorithm with a boundary as the direct product of the 1D algorithm with a boundary and the 1D algorithm in free space.

\subsection{Three dimensions}
\label{sec:bou3D}

We now consider the 3D case. Since we have considered two different 3D LMC algorithms in free space, the ``direct product'' algorithm with simultaneous moves in three directions and the unique optimal algorithm without such moves, we show how to treat a reflective boundary for each of these two algorithms in turn. We assume that the $x=0$ plane serves as the boundary.

\subsubsection{The direct product algorithm}

The extension of the direct product algorithm to the case when a boundary is present is straightforward and can be obtained as the direct product of the 1D algorithm with a boundary and two 1D algorithms (or one 2D algorithm) in free space. This leaves the probabilities of moves into bulk sites unchanged compared to the free space algorithm, but the ``primed'' probabilities for the moves into boundary sites are now
\begin{eqnarray}
p'_{-x}&=&2/27,\\
p'_{\pm y}=p'_{\pm z}&=&5/54,\\
p'_{-x,\pm y}=p'_{-x,\pm z}&=&1/54,\\
p'_{\pm y,\pm z}&=&5/216,\\
p'_{-x,\pm y,\pm z}&=&1/216,\\
p'_0&=&10/27.
\end{eqnarray}

\subsubsection{The algorithm without simultaneous moves in three directions}
In this case, the consideration is similar to 2D. Compared to the 2D case, the master equation has additional terms involving $p'_{\pm z}$, $p'_{-x,\pm z}$, and $p'_{\pm y,\pm z}$. The solution of the continuum equation is obtained from Eq.~(\ref{gensol2Dbou}) simply by introducing the same dependence of the integrand on $k_z$ as its dependence on $k_y$ and integrating over $k_z$ from $-\infty$ to $\infty$. The matching equation (\ref{match2Dbou}) becomes
\begin{eqnarray}
\exp(-a^2(k_x^2+k_y^2+k_z^2)/6)\overset{?}{\simeq}p'_0&+&p'_{+y}e^{-ik_ya}+p'_{-y}e^{ik_ya}+p'_{+z}e^{-ik_za}+p'_{-z}e^{ik_za}\nonumber\\
& &\hspace{-6cm} +p'_{+y,+z}e^{-ik_ya-ik_za}+p'_{+y,-z}e^{-ik_ya+ik_za}+p'_{-y,+z}e^{ik_ya-ik_za}+p'_{-y,-z}e^{ik_ya+ik_za}\nonumber\\
& &\hspace{-6cm} +(p'_{-x}+p'_{-x,+y}e^{-ik_ya}+p'_{-x,-y}e^{ik_ya}+p'_{-x,+z}e^{-ik_za}+p'_{-x,-z}e^{ik_za})\frac{\cos(k_x(d+a))}{\cos(k_xd)}.
\end{eqnarray}
First of all, there are new types of terms in the expansion with no analog in 2D, namely, those involving $y$ and $z$ ($k_yk_z$, $k_y^2k_z$, $k_yk_z^2$, $k_y^2k_z^2$, $k_y^3k_z$, and $k_yk_z^3$). Only four of the resulting equations are independent and they involve only four probabilities (those corresponding to simultaneous moves along $y$ and $z$ axes). Moreover, these equations are the same as they would be for a site in the bulk, since only moves parallel to the boundary are involved. Thus the four probabilities involved are the same as for bulk sites:
\begin{equation}
p'_{+y,+z}=p'_{+y,-z}=p'_{-y,+z}=p'_{-y,-z}=1/36.
\end{equation}
Equations~(\ref{2Dboukx2}) and (\ref{2Dboukx4}) will have the probabilities of the $(-x,\pm z)$ moves added to the sums of the probabilities entering these equations. The normalization condition [Eq.~(\ref{2Dbousum2})] becomes
\begin{eqnarray}
p'_{-x}+p'_{-x,+y}+p'_{-x,-y}+p'_{-x,+z}+p'_{-x,-z}+p_0+p_{+x}+p_{+y}+p_{-y}+p_{+z}+p_{-z}& &\nonumber\\
+p_{+x,+y}+p_{+x,-y}+p_{+x,+z}+p_{+x,-z}+p_{+y,+z}+p_{+y,-z}+p_{-y,+z}+p_{-y,-z}&=&\nonumber\\
p'_{-x}+p'_{-x,+y}+p'_{-x,-y}+p'_{-x,+z}+p'_{-x,-z}+5/6&=&1.\label{3Dbousum2}
\end{eqnarray}
Just as in 2D, all three equations [the modified Eqs.~(\ref{2Dboukx2}) and (\ref{2Dboukx4}) and Eq.~(\ref{3Dbousum2})] involve the same sums of ``primed'' probabilities, which can thus easily be eliminated, and $d=a/2$ follows naturally. Equations~(\ref{2Dboukx2ky}) and (\ref{2Dboukx2ky2}) remain unchanged, and there will also be analogous equations involving the $z$ axis instead of the $y$ axis, so
\begin{equation}
p'_{-x,+y}=p'_{-x,-y}=p'_{-x,+z}=p'_{-x,-z}=1/36.
\end{equation}
Then from Eq.~(\ref{3Dbousum2}),
\begin{equation}
p'_{-x}=1/18.
\end{equation}
Equations~(\ref{2Dbouky}) and (\ref{2Dbouky2}) will have the probabilities of the $(\pm y,\pm z)$ moves added to them (or subtracted in the case of the $(-y,\pm z)$ moves in the first of these equations). From these equations,
\begin{equation}
p'_{+y}=p'_{-y}=1/12.
\end{equation}
Equations obtained by matching $k_z$ and $k_z^2$ terms are analogous to Eqs.~(\ref{2Dbouky}) and (\ref{2Dbouky2}), and from them similarly
\begin{equation}
p'_{+z}=p'_{-z}=1/12.
\end{equation}
Finally, Eq.~(\ref{2Dbouk0}) is modified to include the sum of all ``primed'' probabilities, and from it,
\begin{equation}
p'_0=7/18.
\end{equation}
All other equations (modified as appropriate) and their analogs involving the $z$ axis become identities once the above values are inserted in them.

\subsubsection{Relations for the probabilities}
We note that in both 3D algorithms the probabilities of all allowed diagonal moves and $p'_{-x}$ remain unmodified compared to the bulk. All other probabilities have changed and can be written (again, for both 3D algorithms) as
\begin{eqnarray}
p'_{+y}&=&p_{+y}+p_{-x,+y},\\
p'_{-y}&=&p_{-y}+p_{-x,-y},\\
p'_{+z}&=&p_{+z}+p_{-x,+z},\\
p'_{-z}&=&p_{-z}+p_{-x,-z},\\
p'_0   &=&p_0   +p_{-x}.
\end{eqnarray}
Just as in 2D, this can be interpreted as the diagonal moves across the boundary being projected along the boundary, rather than rejected. Also, as in 1D and 2D, the equivalence of the master equations for boundary and bulk sites once fictitious sites behind the boundary are introduced still holds, again, for both algorithms.

\section{Absorbing boundaries}
\label{sec:bouabs}
Another case of practical interest is that of \textit{absorbing boundaries}. In this case, particles reaching the boundary are ``absorbed'' and disappear from the system. One reason for the importance of this problem is that the FPT problem can be reduced to it: the cumulative FPT distribution is equal to the fraction of absorbed particles as a function of time. In the continuum diffusion problem, the absorbing boundary condition is simply
\begin{equation}
n=0
\end{equation}
at the boundary. In LMC, since the particle number is no longer conserved, there will inevitably be a new type of ``move'' for particles in sites adjacent to the boundary, during which the particle simply disappears. The probability of this occurring does not enter the master equations explicitly (the form of these equations remains unchanged, although the values of the probabilities change), but the probabilities of all other moves will no longer sum up to unity, so the normalization conditions for boundary sites cannot be used (but those for bulk sites are still valid).

\subsection{One dimension}
\label{1Dbouabs}
The general solution of the continuum diffusion equation with an absorbing boundary at $x=0$ is
\begin{equation}
n(x,t)=\int_{0}^{\infty} C(k)\sin(kx)\exp(-Dk^2 t)dk,\label{gensol1Dbouabs}
\end{equation}
i.e., the same as for the reflecting boundary [Eq.~(\ref{gensol1Dbou})], except the cosine is replaced by the sine. In the discrete case, the master equation for the boundary site is still given by Eq.~(\ref{1Dbound}). Plugging the integrand of Eq.~(\ref{gensol1Dbouabs}) into this master equation, replacing $\tau$ with $a^2/6D$ and dividing both parts by $C(k)\sin(kd)\exp(-Dk^2t)$, we get
\begin{equation}
\exp(-k^2a^2/6)\overset{?}{\simeq}p_{0\to 0}+p_{1\to 0}\frac{\sin(k(d+a))}{\sin(kd)}.
\end{equation}
Expanding this up to $O(k^4)$, we obtain
\begin{equation}
1-\frac{k^2a^2}{6}+\frac{k^4a^4}{72}\overset{?}{=}p_{0\to 0}+p_{1\to 0}\frac{d+a}{d}\left(1-\frac{k^2}{6}a(2d+a)+\frac{k^4}{360}(3a^4+12a^3d+8a^2d^2-8ad^3)\right).\label{matching1Dabs}
\end{equation}
It is also convenient (but not necessary) to use the condition that the sum of the probabilities of all moves leaving site 1 is unity (since this site is not adjacent to the boundary). Given that two of these moves ($1\to 1$ and $1\to 2$) lead to a bulk site, their probabilities should coincide with the bulk probabilities ($p_{j\to j}=2/3$ and $p_{j+1\to j}=1/6$, respectively), and therefore the remaining probability, $p_{1\to 0}$, should equal 1/6. Using this in Eq.~(\ref{matching1Dabs}) and equating the $k^2$ terms, we get a quadratic equation for $d$ that has two solutions:
\begin{equation}
d^{(1)}=a;\ d^{(2)}=a/2.\label{d1Dabs}
\end{equation}
Equating the $k^0$ terms and using Eq.~(\ref{d1Dabs}) and $p_{1\to 0}=1/6$, we get
\begin{equation}
p_{0\to 0}^{(1)}=2/3;\ p_{0\to 0}^{(2)}=1/2.
\end{equation}
As mentioned, in both cases
\begin{equation}
p_{1\to 0}^{(1)}=p_{1\to 0}^{(2)}=1/6.
\end{equation}
Both of these solutions turn the equation obtained from equating the $k^4$ terms into an identity; in fact, using that equation instead of the normalization condition would have given the same two solutions. Note that for the boundary site 0 the normalization condition is not satisfied: the sum of the probabilities is $5/6$ for the first solution and $2/3$ for the second one. The complements of these sums to unity ($1/6$ and $1/3$, respectively) can be interpreted as the probabilities of particle disappearance (or absorption by the boundary).

\begin{figure}
\begin{center}
\includegraphics[width=3.5in]{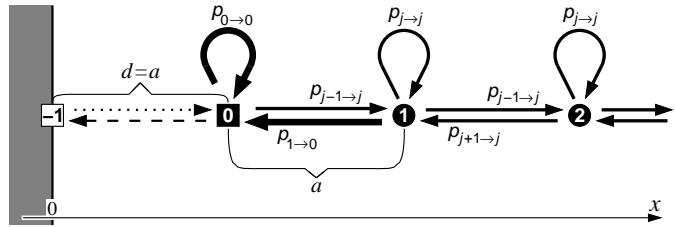}
\caption{One of the two variants of treatment of an absorbing boundary in the 1D optimal algorithm. The inaccessible region is shaded. The site closest to the boundary is the boundary site (filled square); all other sites are bulk sites (filled circles). In this variant, the placement of the boundary site is at the distance $d=a$ from the wall. The open square is a fictitious site at the wall useful in the analysis of the algorithm, as described in the text. The move left from the boundary site (dotted line) leads to the disappearance of the particle. Only the probabilities of the moves to the boundary site (thick arrows) are allowed to change in the analysis, but turn out to be the same as in the bulk for the optimal algorithm. Choosing $d=a/2$, as in Fig.~\ref{fig:bou1D}, is also possible, but leads to different probabilities of the moves.\label{fig:bou1Dabs}}
\end{center}
\end{figure}

Let us now consider the physical interpretation of the two solutions. Note first of all that the first solution corresponds to the absorption probability of $1/6$, which is the same as the probability of the move to the left from a bulk site, and the probabilities of all allowed moves are unchanged compared to the bulk. Therefore, when using the first algorithm, we can treat the boundary site as a bulk site, except whenever a move to the left is attempted, the particle is deleted instead. There is no such simple interpretation for the second solution. Just as for the reflecting boundary, we can also analyze the algorithms by looking at the corresponding master equation. If the first set of parameters is used in the master equation (\ref{1Dbound}), the resulting equation coincides with that for bulk sites [Eq.~(\ref{1Ddiscr})], if a fictitious site (numbered $-1$) is introduced and the mean particle number at that site is made identically equal to zero. The physical interpretation of this is clear: if $d=a$, then site $-1$ is placed right at the absorbing boundary (Fig.~\ref{fig:bou1Dabs}), where the particle concentration in the continuum problem is indeed exactly zero. On the other hand, if the second set of parameters is used in the master equation, the resulting equation will coincide with that for bulk sites, if $n_{-1}=-n_0$. Since for $d=a/2$ the wall is halfway between sites $-1$ and 0, we get $n=0$ at the wall simply by interpolating between $n_0$ and $n_{-1}$. Of course, \textit{linear} interpolation can be done for any $d$; for the interpolated value at $x=0$ to be zero,
\begin{equation}
n_{-1}=-\frac{a-d}{d}n_0.\label{interpol}
\end{equation}
However, it is only at $a=d/2$ and $a=d$ that at least the cubic correction to the interpolated value vanishes and thus the interpolation is more precise (in the latter case, of course, no interpolation is involved, as site $-1$ is located right at the boundary).

Note that if we are content with the linear accuracy of interpolation, we can use Eq.~(\ref{interpol}) in Eq.~(\ref{1Ddiscr}) and by comparison with Eq.~(\ref{1Dbound}) obtain
\begin{equation}
p_{0\to 0}=p_{j\to j}-p_{j-1\to j}\frac{a-d}{a}.
\end{equation}
Note that in the ordinary algorithm, $p_{j\to j}=0$ and so $p_{0\to 0}$ is non-negative and thus physically meaningful only for $a=d$. Thus there is only one way of implementing the absorbing boundary condition within the framework of the ordinary algorithm. Introducing a nonzero probability of staying put broadens the set of allowed values of $d$ and thus gives more freedom. In particular, for $p_{j\to j}=2/3$ and $p_{j-1\to j}=1/6$, as in the optimal algorithm, any $d\ge a/5$ is possible. Even when optimality is required, we still have two choices, $d=a/2$ and $d=a$, more than for the ordinary algorithm. This is yet another advantage of using the optimal algorithm, or nonzero $p_{j\to j}$ in general.

Given the choice between the two optimal algorithms (with $d=a$ and $d=a/2$), which one should be preferred? In principle, the first choice has certain advantages. First, it follows from the previous discussion that the boundary condition $n=0$ is explicit in the first algorithm, but is obtained indirectly via interpolation in the second. It is not obvious, however, that this improves the accuracy of the algorithm overall, as in both cases the continuum solution satisfies the respective master equations to the same order [$O(k^4)$]. Second, recalling the correspondence between diffusion with absorbing boundaries and the FPT problem, we note that it is the variant with $d=a$ that translates directly into the discrete FPT problem as considered in Sec.~\ref{sec:FPT}: in this case the rate of reaching a site is equal to the absorption rate when that site is replaced by an absorbing wall. This means that this variant of the algorithm will produce two correct moments of the absorption time distribution; it can be checked that this is not the case for the other variant. However, there can be other situations when the $d=a/2$ variant is preferable.

\subsection{Two dimensions}
As in the case of a reflecting boundary, we assume that $x=0$ is the boundary and $x>0$ is the accessible region. We can derive the parameters of the optimal algorithm using the same approach as in the reflecting boundary case, by writing down the continuum solution, which in this case is
\begin{equation}
n(x,y,t)=\int_{k_x=0}^{\infty}\int_{k_y=-\infty}^{\infty} C(k_x,k_y)\sin(k_x x)\exp[ik_yy-D(k_x^2+k_y^2)t]dk_x dk_y,\label{gensol2Dbouabs}
\end{equation}
plugging the integrand into the master equation (\ref{master2Dbou}) and demanding that the resulting equality is satisfied as accurately as possible. A much shorter route is obtaining the optimal algorithm as the direct product of the 1D algorithm with an absorbing boundary and another one in free space. Since there are two variants of the optimal 1D algorithm (with $d=a$ and with $d=a/2$), we end up with two variants of the 2D algorithm: variant 1 ---
\begin{eqnarray}
d^{(1)}&=&a,\\
p_0^{'(1)}&=&4/9,\\
p_{-x}^{'(1)}=p'_{\pm y}&=&1/9,\\
p_{-x,\pm y}^{'(1)}&=&1/36,
\end{eqnarray}
and variant 2 ---
\begin{eqnarray}
d^{(1)}&=&a/2,\\
p_0^{'(1)}&=&1/3,\\
p_{-x}^{'(1)}&=&1/9,\\
p_{\pm y}^{'(1)}&=&1/12,\\
p_{-x,\pm y}^{'(1)}&=&1/36.
\end{eqnarray}
The particle disappearance probabilities are $1/6$ and $1/3$, respectively, as in 1D. Note that in the first variant all probabilities are the same as the bulk probabilities. Therefore, as in 1D, the boundary sites can be treated as the bulk sites, except moves into the walls are replaced by particle disappearance. This is not the case for the second variant. Also, it is still true, as in 1D, that the master equations for the boundary sites coincide with those for the bulk sites with $n_{-1,l}=0$ for the first variant and $n_{-1,l}=-n_{0,l}$ for the second variant.

The disadvantage of the direct product approach is that, strictly speaking, we do not prove that the resulting sets of parameters are the only possible ones. A much longer consideration using the first method described above shows that the two variants of the algorithm we have obtained are indeed the only sets of parameters turning the master equation into an identity up to $O(k^4)$ when the continuum solution [Eq.~(\ref{gensol2Dbouabs})] is inserted into it.

\subsection{Three dimensions}

In 3D, both in free space and with a reflecting boundary, we considered two algorithms: the ``direct product'' algorithm and the algorithm with moves in one and two directions only. We now extend these considerations to the case of an absorbing boundary at $x=0$.

\subsubsection{The direct product algorithm}
The direct product algorithm is obtained as the direct product of the 2D algorithm with an absorbing boundary and a 1D algorithm in free space. Since there are two variants of the former, we end up with two variants of the 3D algorithm. In the first variant,
\begin{eqnarray}
d^{(1)}&=&a,\\
p_0^{'(1)}&=&8/27,\\
p_{-x}^{'(1)}=p_{\pm y}^{'(1)}=p_{\pm z}^{'(1)}&=&2/27,\\
p_{-x,\pm y}^{'(1)}=p_{-x,\pm z}^{'(1)}=p_{\pm y,\pm z}^{'(1)}&=&1/54,\\
p_{-x,\pm y,\pm z}^{'(1)}&=&1/216;
\end{eqnarray}
in the second variant,
\begin{eqnarray}
d^{(2)}&=&a/2,\\
p_0^{'(2)}&=&2/9,\\
p_{-x}^{'(2)}&=&2/27,\\
p_{\pm y}^{'(2)}=p_{\pm z}^{'(2)}&=&1/18,\\
p_{-x,\pm y}^{'(2)}=p_{-x,\pm z}^{'(2)}&=&1/54,\\
p_{\pm y,\pm z}^{'(2)}&=&1/72,\\
p_{-x,\pm y,\pm z}^{'(2)}&=&1/216.
\end{eqnarray}

\subsubsection{The algorithm without simultaneous moves in three directions}
To quickly derive this algorithm, we use an approach based on the correspondence between the master equations for the boundary and bulk sites, as described below.

First, we have mentioned when discussing the 1D algorithms in Sec.~\ref{1Dbouabs} that for the variant with $d=a$ these master equations coincide when $n_{-1}$ is put equal to zero in the bulk equation. Extending this to 3D, we replace $n_{-1,l,m}$ with zero for all $l$ and $m$ in the 3D bulk master equation for the evolution of $n_{0,l_0,m_0}$ and compare to the master equation for a boundary site. The result is that all probabilities remain as in the bulk [i.e., are given by Eqs.~(\ref{no3dir1})--(\ref{no3dir3})], except that moves involving the $+x$ direction are, of course, impossible; also, the complement of the sum of the probabilities to unity (which equals $1/6$, as in 1D and 2D) is the probability of the particle disappearance. The same interpretation as in 1D and 2D is still valid: the boundary sites can be treated as bulk sites, but all attempts to move into the boundary are replaced by particle disappearance.

On the other hand, for the variant with $d=a/2$ the 1D master equations for the boundary and bulk sites coincide when $n_{-1}=-n_0$ in the latter. Again, extending this to 3D, we obtain the following probabilities:
\begin{eqnarray}
p'_0=p_0-p_{+x}&=&5/18,\\
p'_{-x}=p_{-x}&=&1/18,\\
p'_{\pm y}=p'_{\pm z}=p_{\pm y}-p_{+x,\pm y}=p_{\pm z}-p_{+x,\pm z}&=&1/36,\\
p'_{-x,\pm y}=p'_{-x,\pm z}=p'_{\pm y,\pm z}=p_{-x,\pm y}=p_{-x,\pm z}=p_{\pm y,\pm z}&=&1/36.
\end{eqnarray}
The sum of all these ``primed'' probabilities is $2/3$, and so the disappearance probability is $1/3$, as in 1D and 2D.

\section{Discussion}
\label{sec:disc}
In this paper, we have obtained the sets of parameters (transition probabilities and the time step) of LMC algorithms for particle diffusion problems that optimize the accuracy of these algorithms. The problem was solved by demanding that the solutions of the master equation of the algorithm approximate those of the continuum diffusion equation as accurately as possible. The matching between the continuum equations and the discrete master equations was done using the general solution of the continuum equation written as a Fourier expansion in space where each component decays exponentially in time. The general solution of the master equation written in a similar form can also be used, in which case the goal is to make the decay rates of the Fourier modes as close as possible in the continuum and discrete equations. However, having the solution of the master equation is not necessary: for example, we did not use it in Secs.~\ref{sec:bourefl} and \ref{sec:bouabs}, where we considered the treatment of boundaries. In that case, the solution of the continuum equation was simply inserted into the master equation and the parameters of the latter adjusted to satisfy the resulting equality as accurately as possible. Having the solution of the master equation, while not necessary, allowed the illustrative comparison between the dispersion relations presented in Fig.~\ref{fig:disp1D}.

Since LMC simulations are generally carried out in order to study long length and long time scale diffusion processes, all expressions are expanded in the wavenumber $k$ and terms of the lowest order are matched. Matching the dispersion relations up to $O(k^2)$ guarantees that the average velocity (equal to zero in the unbiased case) and the diffusion constant are reproduced correctly by the algorithm. In other words, both the mean displacement (the first moment of the distribution for particles starting at a particular site; again, equal to zero) and the mean-square displacement (the second moment) remain correct at all times. Moreover, for all higher-order moments, the leading term at large times has the correct coefficient. However, these higher moments may be incorrect for shorter times. The algorithms can suffer from some artifacts, as discussed in Sec.~\ref{sec:unb1D} for the 1D case. Going to $O(k^4)$, which is possible with only first neighbor moves in 1D and when both first neighbor and second neighbor (diagonal) moves are included in 2D and 3D, removes the artifacts and makes the fourth moment, as well as the \textit{subleading} terms of even higher moments, correct, too, which we have shown for the 1D case, but expect to generalize to 2D and 3D as well. We refer to the resulting algorithms as \textit{optimal}, in the sense that they achieve the best accuracy given the set of moves, although they are not optimal in terms of the simulation speed if accuracy is not a concern. The latter fact is due, in particular, to a distinctive feature of the optimal algorithms: a nonzero probability for a particle not to move during a particular time step, or a \textit{waiting time}.

We have also shown for the 1D case that for a particle starting at a particular point between two boundaries, the optimal algorithm reproduces correctly the first two moments of the distribution of the times of first passage to the boundaries, if both the initial point and the boundaries coincide with lattice sites. In fact, the optimal algorithm can be obtained based on the requirement that the first two moments of the first-passage time are reproduced correctly \textit{or} on the requirement that the first four moments of the particle distribution are correct, instead of matching the dispersion relations. We have also studied the full distributions of the first-passage times comparing them to the exact continuum result. We have shown that the optimal algorithm converges much faster to the exact result than the ordinary algorithm without a waiting time as the mesh step decreases, so if a particular accuracy is required, a much coarser mesh can be used with the optimal algorithm, which provides a speedup that in the end more than compensates for the loss of computational speed due to the waiting time.

One fact worth mentioning is an analogy between the optimal LMC algorithms and \textit{Lattice Boltzmann} (LB) algorithms for mesoscopic fluid simulations~\cite{LBreview,dunweg08}.  LB equations can be thought of as master equations for particles residing at the sites of a lattice and moving at each time step to a predetermined set of nearby sites, as in LMC. However, a crucial difference is that the probabilities of particular moves are not fixed. At each site, the mean numbers of particles with each particular \textit{velocity} are considered separately and evolve individually according to the LB equations. The velocity of a particle determines how that particle moves at the next time step, and since the set of moves is discrete, the set of possible velocities in the algorithm is discrete as well. In different varieties of the LB approach different sets of discrete velocities (and thus of possible moves) are used. The standard notation used to distinguish these varieties is D$m$Q$n$, where $m$ is the dimensionality of the space and $n$ is the number of possible velocity vectors. One question relevant to the design of LB algorithms is finding the best approximation of the equilibrium Maxwell velocity distribution using a particular set of allowed discrete velocity vectors. It turns out that in the best approximation, the mean numbers of particles with particular velocities are proportional to the probabilities of the corresponding moves in the optimal LMC algorithm with the same set of moves. For the D2Q9 LB algorithm~\cite{LBchinese}, this will be our 2D optimal LMC; for the D3Q19 algorithm~\cite{dunweg08}, our 3D LMC with moves along one and two directions; for the D3Q27 algorithm~\cite{LBchinese}, our 3D ``direct product'' algorithm. This correspondence is not coincidental. The Maxwell distribution is a Gaussian distribution, as is the distribution of positions of particles diffusing in continuum space at a given time after they started at the same point. Given this, it is easy to see that the problem of finding the distribution of discrete velocities approximating the Maxwell distribution is mathematically equivalent to finding the set of probabilities of moves such that after a single step starting from the same site the resulting particle distribution is as close to Gaussian as possible. Of course, even though there is a mathematical equivalence, the physical meaning of the problems is very different.

Another interesting result of this paper concerns the treatment of impenetrable reflecting boundaries. Often in MC algorithms, if an attempted move is forbidden, e.g., because a particle would overlap with an obstacle, it is simply rejected. This is a correct choice in algorithms that only intend to reproduce equilibrium properties, as it preserves detailed balance, which is all that matters in that case. However, as we have shown, this is not the best choice in a \textit{dynamical} algorithm in a two- or higher-dimensional space. In that case, the best accuracy is achieved by replacing forbidden moves across the boundary with their projections along that boundary. This result makes physical sense: if, for example, in 2D all boundaries are orthogonal to the $x$ direction and are infinitely long in the $y$ direction, the diffusion in the $y$ direction should not be affected, which is only possible if the simultaneous moves in the $x$ and $y$ directions are projected along the $y$ direction.

We have also considered the case of absorbing boundaries. It is interesting that for each of the optimal free space algorithms, there are two ways of treating the boundaries parallel to one of the lattice axes. The more obvious one is to destroy the particle whenever it moves into the wall, but leave all other moves unmodified. This is the best approximation for a wall that is at the distance equal to the mesh step from the last row of sites next to the boundary ($d=a$ using the notation of the paper). However, there is also a second variant, with rules that do not have a simple interpretation, which corresponds to the last row of sites being at the distance from the boundary of half the mesh step ($d=a/2$). This second variant (as well as intermediate, non-optimal choices) is only possible in algorithms with a waiting time (even if optimality is not required).

\begin{figure}
\begin{center}
\includegraphics[width=3.5in]{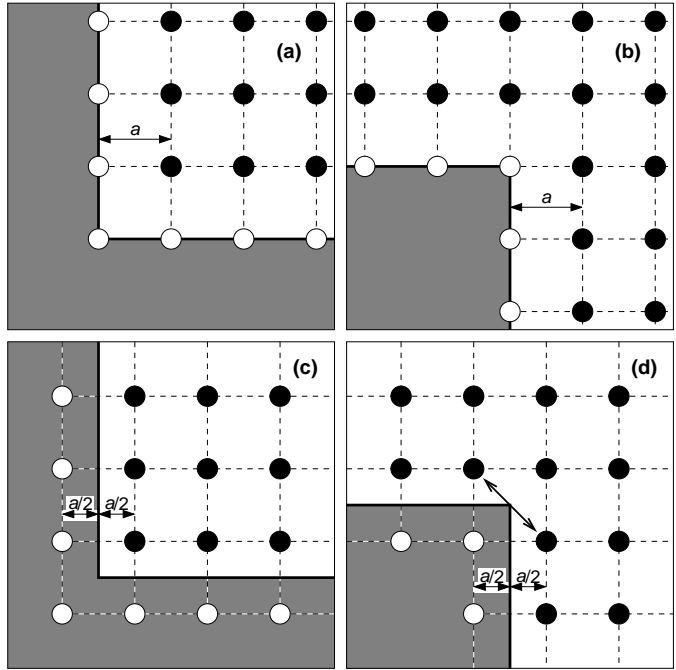}
\caption{2D examples of different configurations of walls forming corners and different variants of meshes for optimal LMC algorithms. Forbidden areas are shaded; real sites are black circles and fictitious sites introduced to facilitate the analysis of the algorithms are white circles. Cases (a), (b) and (c) can be analyzed, as described in the text, but in the case (d) there are complications that physically correspond to the uncertainty about the probabilities of the diagonal transitions ``across the corner'' indicated by the arrow.\label{fig:corners}}
\end{center}
\end{figure}

Note that we have only treated the case of infinite, flat boundaries. In real situations of interest, one deals with finite and/or curved boundaries (finite obstacles, tortuous pores, etc.) In the simplest case, the walls consist of flat pieces, each of which is orthogonal to one of the axes, that are joined at corners, like those in Fig.~\ref{fig:corners}. An example would be a rectangular obstacle in 2D with the sides along the axes. Sites adjacent to the walls that are far from the corners can be treated as boundary sites next to infinite walls (this assumes that it is possible to choose the lattice so that the distance $d$ from the boundary sites to the boundary is as required by the algorithm everywhere). However, we still need a way to treat sites next to the corners. We note, though, that we have assumed throughout that the particle concentration varies smoothly on the length scale of the mesh step, and this implies that any features of the obstacles and the walls are likewise smooth on this length scale, so corners should be rare. It does not matter much then how exactly these rare situations are treated; we will consider these complications in a separate paper in the future. However, we can offer some intuitive guidance in some of these situations. For example, it is intuitively clear how to deal with absorbing boundaries in the algorithms with $d=a$ [Fig.~\ref{fig:corners}, (a) and (b)]: in these algorithms, all moves into the wall lead to absorption and all other moves are unchanged, which generalizes straightforwardly to corners. It is also clear how to handle the situation with $d=a/2$ (either absorbing or reflecting walls), when, e.g., in 2D the particle is allowed to move in one quadrant and the other three are blocked [Fig.~\ref{fig:corners} (c)]. In this case, fictitious sites are introduced inside the forbidden area symmetrically with respect to the boundary to the row of sites nearest to it [Fig.~\ref{fig:corners} (c)], and the mean particle numbers at these sites are assumed equal to those at sites nearest to them across the boundary (with the minus sign in the case of absorption). The generalization to the case of a corner is straightforward, by assuming that the fictitious site closest to the corner takes on the value of the real site closest to the corner (possibly with the minus sign). However, in the otherwise analogous case when three quadrants are free and one is blocked [Fig.~\ref{fig:corners} (d)], there is some ambiguity, as it is not clear which of the three real sites closest to the corner should give its value to the fictitious site nearest the corner. This translates into an uncertainty about the probabilities of the diagonal moves ``passing through the corner'' shown in the figure.

Curved boundaries or those not orthogonal to the axes are even more complicated. Approximating such boundaries by piecewise flat boundaries with each piece orthogonal to one of the axes, in such a way that those sites inside the original boundary are still inside and those outside are still outside, would in effect produce ``jagged'' boundaries with many corners adjacent to each other that need special treatment. A more accurate approach would be approximating such boundaries with flat pieces oriented \textit{arbitrarily}. However, this also requires a separate consideration. The rules we have derived cannot be extended straightforwardly to such ``skewed'' boundaries, even when they are perfectly flat: for instance, projecting forbidden moves along the boundary is impossible, as such a projection would not coincide with any of the allowed moves. Solving these problems will be a subject of future work. Considering finite hard particles of an arbitrary shape is a more complicated problem, since particle rotation needs to be taken into account (spherical particles are an exception). An even more complicated problem is that of several diffusing hard-core particles that can collide with each other. Such particles would move independently between collisions (according to the rules derived in this paper, if rotation can be ignored), but rules for treating collisions still need to be worked out.

The preceding discussion may sound as if there is still some way to go before the algorithms described here become practically useful. This is not entirely true, however. As mentioned in Sec.~\ref{sec:compcont}, our 1D optimal algorithm is the zero-field limit of a previously derived algorithm for biased diffusion~\cite{gauthier04}. We have recently applied the latter algorithm to a 1D model of polymer translocation through a nanopore~\cite{gauthier08} and showed that this algorithm gives very different widths of the translocation time distributions compared to the ordinary algorithm without waiting times, thus demonstrating the importance of introducing waiting times~\cite{transloc10}. It is true, though, that the utility of the new algorithms will be further enhanced when the issues described above are taken care of, as well as when the approach is extended to biased diffusion in 2D and 3D as well. In general, we expect the new algorithms to be useful in those problems where the transient behavior of the system at short times is important. This includes various first-passage problems (including the polymer translocation problem mentioned above), as well as adsorption, aggregation~\cite{fingerDLA}, anomalous diffusion~\cite{langowski} and reaction-diffusion~\cite{bernstein} problems. One example of a problem dealing with transient effects where our approach may be useful is studying drug release profiles from disordered porous matrices~\cite{casault07,kosmidis}. On the other hand, if only the long-time behavior is of interest and short-time transient dynamics does not affect it, then the ordinary algorithms with jumps at every time step may be just as good and may offer computational time savings compared to the optimal algorithms.

In those cases where discrete-time LMC can be used, there is also an option to use MC algorithms with continuous space (off-lattice MC), continuous time (often referred to as kinetic MC~\cite{srolovitz,voter,KMC2007,sinnoKMC09}) or both. Of course, we do not claim that LMC is always the best option; rather, the goal of this paper is to suggest that in those cases where the choice in favor of LMC is made, one might consider improving the accuracy by using the LMC modifications described here. We do note, however, that since the choice between LMC and continuous MC is often driven by the tradeoff between higher speed of the former and higher accuracy of the latter, improving the accuracy of LMC may actually make LMC preferable in some situations where this would not be the case otherwise.

\section*{Acknowledgements}

This research was supported by grants from NSERC and the NIH (Grant No. 2 R01 HG001970-07 through Stanford University). The findings, opinions and recommendations expressed in this letter are those of the authors and not necessarily those of Stanford University or the NIH.

\section*{Appendix: a derivation of the long-time dependence of the first-passage rate for the LMC algorithms}
\setcounter{equation}{0}
\renewcommand{\theequation}{A.\arabic{equation}}
The 1D first-passage problem for LMC algorithms, as formulated in Sec.~\ref{sec:FPT}, can be solved by  considering the master equations (\ref{1Ddiscr}) for sites from $-N+1$ to $N-1$ and fixing
\begin{equation}
n_{\pm N}=0.\label{bcFPT}
\end{equation}
The solution of this system will give the mean numbers of particles at every site that have not yet reached the walls at sites $\pm N$. This solution can be obtained from the general solution for the infinite lattice [Eq.~(\ref{gensol1D})] by picking those modes that satisfy the boundary conditions [Eq.~(\ref{bcFPT})], which gives $k_m=\pi (2m+1)/2Na$ with $m=0,\ldots,N-1$ and $C(-k)=C(k)$; thus the solution is
\begin{equation}
n_j(t)=\sum_{m=0}^{N-1} C_m \cos(\pi (2m+1)j/2N)\exp [-\alpha_d(k=\pi (2m+1)/2Na)t].\label{genint}
\end{equation}
The mean number of particles reaching the walls at time step occurring at time $t$, $\nu(t)$, is proportional to the number of particles at adjacent sites at the previous time step, i.e.,
\begin{equation}
\nu(t)=p_{j-1\to j}n_{N-1}(t-\tau)+p_{j+1\to j}n_{-N+1}(t-\tau).
\end{equation}
This number per unit time (i.e., the rate) is
\begin{equation}
r(t)=\frac{\nu(t)}{\tau}=\frac{2D}{a^2}n_{N-1}(t-\tau),\label{FPratedis}
\end{equation}
where we have used Eq.~(\ref{p1D}) and the fact that, according to Eq.~(\ref{genint}) or simply from symmetry considerations, $n_{-N+1}=n_{N-1}$.

For the optimal algorithm, $\alpha_d(k)$ is a monotonically increasing function, thus in the large $t$ limit, only the first term in the sum survives, which gives
\begin{equation}
n_j(t)\simeq C_0\cos(\pi j/2N)\exp[-\alpha_d(k=\pi/2Na)t].
\end{equation}
Then from Eq.~(\ref{FPratedis}), 
\begin{equation}
r(t)\simeq \frac{2DC_0}{a^2}\sin(\pi/2N)\exp[\alpha_d(k=\pi/2Na)\tau]\exp[-\alpha_d(k=\pi/2Na)t].
\end{equation}
Therefore the decay rate is
\begin{equation}
\beta_{\rm opt}=\alpha_d(k=\pi/2Na)=-\frac{6DN^2}{b^2}\ln[2/3+(1/3)\cos(\pi/2N)],
\end{equation}
where we have used $a=b/N$ and Eq.~(\ref{alpha1Ddiscr}) with $p$ and $\tau$ given by Eqs.~(\ref{1Dopt1})--(\ref{1Dopt3}). The prefactor is
\begin{equation}
\gamma_{\rm opt}=\frac{2DC_0}{a^2}\sin(\pi/2N)\exp[\alpha_d(k=\pi/2Na)\tau].
\end{equation}
To find $C_0$, we need to expand the initial condition, $n_j(0)=\delta_{j,0}$ in the Fourier series,
\begin{equation}
\delta_{j,0}=\sum_{m=0}^{N-1} C_m \cos(\pi (2m+1)j/2N).
\end{equation}
Since
\begin{equation}
\sum_{j=-N}^{N}\cos(\pi (2m_1+1)j/2N)\cos(\pi (2m_2+1)j/2N)=N\delta_{m_1,m_2},
\end{equation}
we get
\begin{equation}
C_m=\frac{1}{N}\sum_{j=-N}^{N}\delta_{j,0}\cos(\pi (2m+1)j/2N)=\frac{1}{N}\label{Cm}
\end{equation}
for all $m$, including 0, and therefore
\begin{equation}
\gamma_{\rm opt}=\frac{2D}{Na^2}\sin(\pi/2N)\exp[\alpha_d(k=\pi/2Na)\tau]=\frac{2DN\sin(\pi/2N)}{b^2[2/3+(1/3)\cos(\pi/2N)]}.
\end{equation}

For the ordinary algorithm, the situation is slightly more complicated, because $\alpha_d(k)$ is no longer monotonic; as a result, $|\alpha_d(k_{N-1})|=\alpha_d(k_0)$ and two terms of equal magnitude survive at large time:
\begin{eqnarray}
n_j(t)&\simeq& C_0\cos\left(\frac{\pi j}{2N}\right)\exp\left[-\alpha_d\left(k=\frac{\pi}{2Na}\right)t\right]\nonumber\\
& &+C_{N-1}\cos\left(\frac{\pi (2N-1)j}{2N}\right)\exp\left[-\alpha_d\left(k=\frac{\pi(2N-1)}{2Na}\right)t\right].
\end{eqnarray}
The first-passage rate is
\begin{eqnarray}
r(t)&\simeq &\frac{2D}{Na^2}\sin\left(\frac{\pi}{2N}\right)\left\{\exp\left[-\alpha_d\left(k=\frac{\pi}{2Na}\right)(t-\tau)\right]\right.\nonumber\\
& &\left.+(-1)^{N+1}\exp\left[-\alpha_d\left(k=\frac{\pi(2N-1)}{2Na}\right)(t-\tau)\right]\right\},\label{rateord0}
\end{eqnarray}
where we have used Eq.~(\ref{Cm}) for $C_m$ (that is still valid for the ordinary algorithm).
According to Eq.~(\ref{alpha1Ddiscr}) with $p$ and $\tau$ given by Eqs.~(\ref{1Dordinary1})--(\ref{tau1D}),
\begin{eqnarray}
\tau\alpha_d\left(k=\frac{\pi}{2Na}\right)&=&-\ln\cos\left(\frac{\pi}{2N}\right),\\
\tau\alpha_d\left(k=\frac{\pi(2N-1)}{2Na}\right)&=&-\ln\left[-\cos\left(\frac{\pi}{2N}\right)\right]=\tau\alpha_d\left(k=\frac{\pi}{2Na}\right)-i\pi,
\end{eqnarray}
so Eq.~(\ref{rateord0}) becomes
\begin{equation}
r(t)\simeq \frac{2D}{Na^2}\sin\left(\frac{\pi}{2N}\right)\exp\left[-\alpha_d\left(k=\frac{\pi}{2Na}\right)(t-\tau)\right]\times\left[1+(-1)^{N+M}\right],
\end{equation}
where $M$ is the number of the time step ($t=M\tau$). The expression in the rightmost square brackets alternates between 0 and 2, which represents a familiar feature of the ordinary algorithm: no particles reach the boundaries at even steps for odd $N$ and at odd steps for even $N$. Since we have decided to ignore these oscillations, we simply replace this expression with its average value of unity. Then we immediately obtain for the decay rate,
\begin{eqnarray}
\beta_{\rm ord}=\alpha_d\left(k=\frac{\pi}{2Na}\right)&=&-\frac{2DN^2}{b^2}\ln\cos\left(\frac{\pi}{2N}\right),\\
\gamma_{\rm ord}&=&\frac{2DN}{b^2}\tan\left(\frac{\pi}{2N}\right).
\end{eqnarray}

\end{document}